\documentclass[aps,prl,showpacs,twocolumn,superscriptaddress,floats,floatfix, longbibliography]{revtex4-1}
\usepackage{amssymb,amsmath,bm}
\usepackage{graphics}
\usepackage{color}
\usepackage{subfigure}
\usepackage{ulem}
\usepackage{rotating}
\usepackage{hyperref}
\usepackage{nameref}

\newcommand{\VW}{{V_{\rm W}}}

\newcommand{\yPi}{{y_{\rm P0}}}

\newcommand{\ReP}{{Re_{p}}}

\newcommand{\Res}{{Re_{\rm s}}}
\newcommand{\Ls}{{L_{\rm s}}}
\newcommand{\DP}{{d_{\rm p}}}
\newcommand{\Lg}{{L_{\gamma}}}
\newcommand{\lamg}{{\Lambda_{\gamma}}}

\newcommand{\equeqref}[1]{Eq.~\eqref{#1}}
\newcommand{\Fig}[1]{Fig.~(\ref{#1})}
\newcommand{\subfig}[2]{Fig.~(\ref{#1}#2)}

\begin{document}
\title{Conditional stability of particle alignment in finite-Reynolds-number channel flow}
\author{Anupam Gupta}
\email{anupam.gupta@ensiacet.fr}
\altaffiliation{present address: Paulson School of Engineering and Applied Sciences, Harvard University, Cambridge, Massachusetts 02138, USA}
\affiliation{Laboratoire de G\'{e}nie Chimique, Universit\'{e} de Toulouse, CNRS, Toulouse FRANCE}
\affiliation{FERMaT, Universit\'e de Toulouse, CNRS, Toulouse, France}
\author{Pascale Magaud}
\affiliation{Institut Cl\'ement Ader (ICA), Universit\'e de Toulouse, Toulouse, France}
\affiliation{Universit\'e de Limoges, 33 rue Fran\c{c}ois Mitterrand, 87032 Limoges, France}
\author{Christine Lafforgue}
\affiliation{Laboratoire  d'Ing\'enierie des Syst\`emes Biologiques et des Proc\'ed\'es, Universit\'e de Toulouse, CNRS, INRA Toulouse, France}
\author{Micheline Abbas}
\email{micheline.abbas@ensiacet.fr (corresponding author)}
\affiliation{Laboratoire de G\'{e}nie Chimique, Universit\'{e} de Toulouse, CNRS, Toulouse FRANCE}

\begin{abstract}
Finite-size neutrally buoyant particles in a channel flow are known to accumulate at specific equilibrium positions or spots in the channel cross-section if the flow inertia is finite at the particle scale. Experiments in different conduit geometries have shown that while reaching equilibrium locations, particles tend also to align regularly in the streamwise direction. In this paper, the Force Coupling Method was used to numerically investigate the inertia-induced particle alignment, using square channel geometry. The method was first shown to be suitable to capture the quasi-steady lift force that leads to particle cross-streamline migration in channel flow. Then the particle alignment in the flow direction was investigated by calculating the particle relative trajectories as a function of flow inertia and of the ratio between the particle size and channel hydraulic diameter. The flow streamlines were examined around the freely rotating particles at equilibrium, revealing stable small-scale vortices between aligned particles. The streamwise inter-particle spacing between aligned particles at equilibrium was calculated and compared to available experimental data in square channel flow (Gao {\it et al.} Microfluidics and Nanofluidics {\bf 21}, 154 (2017)). The new result highlighted by our numerical simulations is that the inter-particle spacing is unconditionally stable only for a limited number of aligned particles in a single train, the threshold number being dependent on the confinement (particle-to-channel size ratio) and on the Reynolds number. For instance, when the particle Reynolds number is $\approx1$ and the particle-to-channel height size ratio is $\approx0.1$, the maximum number of stable aligned particles per train is equal to 3. This agrees with statistics realized on the experiments of (Gao {\it et al.} Microfluidics and Nanofluidics {\bf 21}, 154 (2017)). It is argued that when several particles are hydrodynamically connected moving as a unique structure (the train) with a steady streamwise velocity, large-scale hydrodynamic perturbations induced at the train scale prohibit small-scale vortex connection between the leading and second particles, forcing the leading particle to leave the train.
\end{abstract}

\maketitle

\section{1. Introduction}

The experiments of Segre and Silberberg~\cite{segre1962behaviour} shed the light on the fact that neutrally buoyant particles experience cross-streamline migration in a parabolic flow if the flow inertia is finite at the particle scale. The dipole interaction with the quadratic part of the flow is responsible of the particle migration. Theoretical computation of the resulting lift force and its dependence on the flow inertia has progressed slowly over decades \cite{ho1974inertial,vasseur1976lateral,schonberg1989inertial,asmolov1999inertial}. Understanding this phenomenon opened a new field of applications with the development of microfluidics, where separation or detection of microparticles is operated by hydrodynamic focusing like flow cytometry~\cite{oakey2010particle}, single cell encapsulation~\cite{edd2008controlled} and cell diagnostics~\cite{hur2010sheathless}. It can be especially practical in the sense that external fields (like electrical, magnetical) or membranes are avoided. \\

\begin{figure} [h!]
\hspace{-0.5cm}
\begin{center}
{\includegraphics[width=0.6\linewidth]{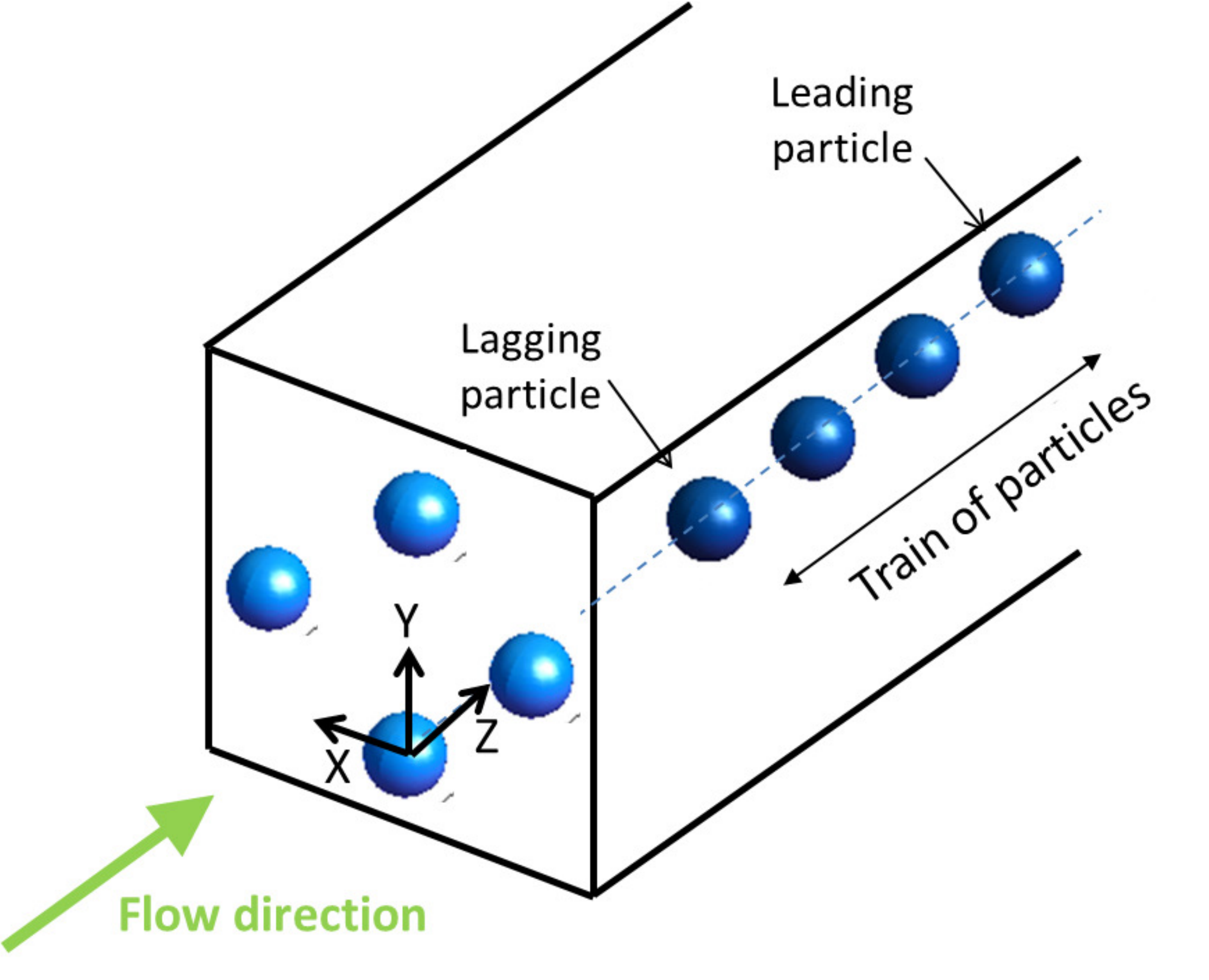}}
\end{center}
\vspace{-0.6cm}
\caption{Scheme showing the possible focusing positions of particles on four spots in the cross-section of a square channel (light blue spheres) and their alignment in the streamwise direction (darker blue). $Z$ indicates the flow direction. $X$ and $Y$ are the directions parallel to the walls. In the simulations particles are placed in the symmetry plane with respect to the $X$ direction. }\label{fig:sketch}
\end{figure}

In more recent experiments, particles were found to accumulate preferentially at equilibrium positions that depend on the conduit cross-section. The accumulation region consists of a ring in a tube flow, and of spots at the center of channel faces in square or rectangular ducts as recently reported ~\cite{matas2004,di2007continuous,bhagat2008enhanced,choi2011lateral}. It has been also observed that, in addition to the existence of equilibrium positions in the cross-section, particles tend to become ordered or evenly spaced in the streamwise direction (so-called trains are formed) \cite{matas2004trains,chun2006inertial,hur2010sheathless,loisel2015inertia,kahkeshani2016preferred,gao2017self}. These observations were obtained in several flow geometries. A sketch of particles assembled in the form of a streamwise train is illustrated in figure \ref{fig:sketch} in the case of square channel flow. \\

These particle assemblies originate from the interaction, in shear flow, of particle pairs at finite flow inertia in the presence of the walls. The experimental observations (usually by optical techniques) of particle trains suggest that at the end of pair interactions, an equilibrium inter-particle (streamwise) spacing is reached. This spacing varies like $Re_p^{-1/2}$ ($Re_p$ being the particle Reynolds number defined at the end of the introduction), as it was obtained in tube and later in square channel flows \cite{matas2004trains},\cite{gao2017self}. Neutrally buoyant particles transported by shear flow induce local streamline reversal at finite inertia \cite{poe1975closed}.  As the inter-particle spacing in the train structures decreases with the flow inertia, it was first suggested by Matas et al. \cite{matas2004trains} that the train formation is related to the flow induced by one particle in finite-inertia shear, as a particle causes the reversal of streamline direction, but a second particle following such a streamline is cut off from receding by the wall. The 2D pair dynamics was later investigated by Yan et al. \cite{yan2007hydrodynamic} in wall-bounded shear (linear) flow. The authors revealed that the particle pair can reach a stable equilibrium or limit cycles at finite inertia, depending on the streamwise boundary conditions. \\

Nevertheless Lee et al \cite{lee2010dynamic} have measured inter-particle spacings in channel flow, at different downstream positions of the channel and plotted histograms. Interestingly, the peak in inter-particle spacing seemed to continuously shift to larger distances further downstream. The authors noted that this shift becomes noticeable after particles travel long distances of order hundred times the channel height, and attributed this to residual viscous repulsive interactions. \\

We show in this paper that particles assembled in the streamwise direction due to finite flow inertia reach stable inter-particle spacings if a small number of particles is involved. However the apparently long-lived trains become unstable if a large number of particles are aligned, in which case the leading particle leaves the train. The corresponding dynamics seems to be very slow. This observation is made possible by simulating the full dynamics of a few particles aligned along the flow direction on a single spot in the square channel, and very long simulation domains to avoid the effect of periodic boundaries in the flow direction or the distant interaction between different trains at different spots. We also show that the maximum size of a stable train depends on the operating conditions that can be gathered under two dimensionless numbers: the particle confinement and the Reynolds number. The particle confinement is defined as the ratio between the particle  diameter $d_p$ and the channel hydraulic diameter $H$. The Reynolds number describes the competition between inertial and viscous forces, either at the channel scale, $Re=UH/\nu$ (the so-called channel Reynolds number), or at the particle scale $Re_p=Re*(d_p/H)^2$ (particle Reynolds number).  $U$ is the average channel flow velocity and $\nu$ is the kinematic viscosity.

The paper is organized as following. The numerical method is described and validated in sections 2 and 3. These two sections are included in order to assess the relevance of the Force Coupling Method for the simulation of inertia-induced particle migration and alignment. The reader can skip these two sections if not interested in numerical details. In section 4, stable particle assemblies are investigated close to equilibrium. The train formation process and the stable train properties are described as a function of the Reynolds number and of the number of aligned particles. Instability of particle alignment is observed as soon as a large number of particles are aligned in the flow direction. The paper is ended with a discussion (section 5) on the possible driving mechanism. \\

Note that regarding the particle size, for notation convenience and comparison with other theoretical and experimental frameworks, the particle radius $a$ is used in section 2, in some places in section 3 and in the appendix. The particle diameter $d_p$ is exclusively used starting from section 4. \\

\section{2. Numerical method formulation}\label{sec:Num}

The description of the numerical method can be found in \cite{wang2017modulation}. It is re-written in this paper before the validation section, for the sake of completeness. Direct numerical simulations of single-phase flows are performed by using the code JADIM for an incompressible Newtonian fluid \cite{calmet1997large}. The unsteady 3-D Navier-Stokes equations discretized on a staggered grid are integrated in space using the finite volume method. All terms involved in the balance equations are written in a conservative form and are discretized using second-order centered schemes in space. The solution is advanced in time by a second-order semi-implicit Runge-Kutta/Crank Nicholson time stepping procedure, and incompressibility is achieved by correcting the pressure contribution which is the solution of the Poisson equation.\\

Numerical simulations of particle trajectories and suspension flow dynamics are based on multipole expansion of momentum source terms added to the Navier-Stokes equations (namely Force-Coupling Method as formulated in \cite{maxey2001localized,lomholt2003force}). Flow equations are dynamically coupled to Lagrangian tracking of particles. The fluid is assumed to fill the entire simulation domain, including the particle volume. The fluid velocity and pressure fields are solutions of continuity \equeqref{eq:FCM1} and momentum balance \equeqref{eq:FCM2} and \equeqref{eq:FCM3}.

\begin{equation} \label{eq:FCM1}
\nabla\cdot{\bm{{u}}}=0
\end{equation} 
\begin{equation} \label{eq:FCM2}
\rho \frac{D\bm{u}}{D\bm{t}}=-\nabla p+\mu \nabla^2\bm{u}+\bm{f}(\bm{x},t)
\end{equation} 
\begin{align} \label{eq:FCM3}
f_i(\bm{x},t)={}&\sum_{n=1}^{N_p}F_i^{n}\triangle(\bm{x}-\bm{Y}^{n}(t))\nonumber\\
&+G_{ij}^{n} \frac{\partial}{\partial x_j}\triangle^\prime(\bm{x}-\bm{Y}^{n}(t))
\end{align} 

$\bm{u}$ is the fluid velocity. $\rho$ and $\mu$ are, respectively, the density and dynamic viscosity of the fluid. The body force distribution $\bm{f}(\bm{x},t)$ in the momentum balance \equeqref{eq:FCM3} accounts for the presence of particles in the flow. It is written as a multipole expansion truncated after the second term. The first term of the expansion called the monopole represents the force $\bm{F}^n$ that the particle exerts on the fluid, due to particle inertia,  external forcing or particle-to-particle contact forces (\equeqref{eq:FCM4}). The second term, called dipole, is based on a tensor $\bm{G}^n$ sum of two contributions: an anti-symmetric part is related to external torques applied on the particle, and a symmetric part that accounts for the resistance of a rigid particle to deformation by ensuring zero average strain-rate inside the particle volume, \equeqref{eq:FCM5}. \\

\begin{equation} \label{eq:FCM4}
\bm{F}^n=(m_p-m_f) \left( \bm{g} - \frac{d\bm{V}^n}{dt} \right) + \bm{F}^n _{ext}
\end{equation}

\begin{equation} \label{eq:FCM5}
S_{ij}^n(t)=\frac{1}{2}\int(\frac{\partial u_i}{\partial x_j}+\frac{\partial u_j}{\partial x_i})\triangle^\prime(\bm{x}-\bm{Y}^{n}(t))d^3x=0
\end{equation} 

$m_p$ and $m_f$ are respectively the mass of the particle and that of the fluid in the region occupied by the particle. 
The particle finite-size is accounted for by spreading the momentum source terms around the particle center $\bm{Y}^{n}$ using a Gaussian spherical envelope, one for the monopole $\triangle(\bm{x})=(2\pi\sigma ^2)^{-3/2} e^{(- \left| \bm{x} \right| /2 \sigma^2)}$, and another one for the dipole  $\triangle^\prime(\bm{x})=(2\pi{\sigma^\prime}^2)^{-3/2}e^{(- \left| \bm{x} \right| /2{\sigma^\prime}^2)}$. The widths of the Gaussian envelopes, $\sigma$ and $\sigma^\prime$ are set with respect to the particle radius $a$ such that the settling velocity and the hydrodynamic perturbation generated by a particle in a shear flow are
both exactly matched to Stokes solutions ($\sigma=a/\sqrt{\pi}$ and $\sigma^\prime=a/(6\sqrt{\pi})^{1/3}$) for a single particle.\\

The particle translation and rotation velocities  are obtained from a local weighted average of the volumetric fluid velocity (resp. rotational velocity) field over the region occupied by the particle (\equeqref{eq:FCM6} and \equeqref{eq:FCM7}).

\begin{equation}\label{eq:FCM6}
\bm{V}^n(t)=\int\bm{u}(\bm{x},t)\triangle(\bm{x}-\bm{Y}^{n}(t))d^3x
\end{equation}
\begin{equation}\label{eq:FCM7}
\Omega ^n(t)=\frac{1}{2}\int(\nabla\times\bm{u}(\bm{x},t))\triangle^\prime(\bm{x}-\bm{Y}^{n}(t))d^3x
\end{equation} 

Particle trajectories are then obtained from numerical integration of the equation of motion as in \equeqref{eq:FCM8}.
\begin{equation}\label{eq:FCM8}
\frac{d\bm{Y}^n}{dt} = \bm{V}^n
\end{equation}

This modelling approach allows calculating the hydrodynamic interactions with a moderate computational cost. For a neutrally buoyant particle, the monople and the anti-symmetric contribution to the dipole are stictly zero. Only the symmetric part of the dipole (Stresslet) allows to account for the interaction between the particle and the shear flow. Eight grid points per particle diameter are usually sufficient to correctly capture this interaction. \\

The method has been validated in the limit of vanishing particle Reynolds number \cite{maxey2001localized,lomholt2003force}. It has later been extended to the case of finite flow inertia at the particle scale, i.e. $Re_p=O(1)$, \cite{yeo2013dynamics,yeo2010modulation}. \citet{loisel2015inertia} have shown that the Stresslet components of a single particle placed in a linear flow compare very well with DNS measurements, up to particle Reynolds number equal to 5 \cite{mikulencak2004stationary}. Additional validation tests with a single particle in quadratic flow are presented in the next section and in appendix A. \\

As for the interaction between two spheres in a linear flow, \citet{yeo2013dynamics} have shown that the FCM gives the right relative particle trajectories at $Re_p=O(1)$. When two particles are initially placed in the shear plane (perpendicular to the vorticity direction), their relative trajectory remains in-plane, and it is open or reversed depending on the initial shift in the shear direction ($\delta y$), of the lagging particle with respect to the leading one. The bifurcation between the two types of trajectories is close to the one found in LBM simulations \cite{Haddadi2015}. The off-plane spiraling interaction is less well captured when the gap between particle surfaces is smaller than $0.1d_p$, however the amplitude of the relative velocity is very small in that case, and it does not play a significant role in the system studied in this paper (particles in the same shear plane).\\

%%%%%%%%%%%%%%%%%%%%%%%%%%%%%%%%%%%%%%%%%%%%%%%%%%%%%%%%%%%%%%%%%%%%%%%%%%%%%%
\section{3. Validation of the numerical method}

At very low Reynolds number, a small neutrally-buoyant spherical particle follows the flow streamlines. Near a wall, both the translational and rotational particle velocities are smaller than the local fluid flow velocities \citep{goldman1967slow}. However, the particle does not experience a wall-normal motion for reversibility reasons. If the flow is slightly inertial at the particle scale, the neutrally buoyant particle experiences lift perpendicular to the flow streamlines, in the presence of shear, the intensity and direction of the lift depending on the flow configuration, and on whether the particle is free to rotate or not. In channel flow, the interaction of the particle Stresslet with the curved background flow profile induces a lift force oriented toward the channel walls when the particle is located near the central region \cite{ho1974inertial}. This force is enhanced by flow inertia. When the particle is very close to the wall, the particle slip is large. The particle slip in the presence of shear near a wall leads to a lift force oriented toward the high velocity region (as computed for instance by \citet{cherukat1994inertial}) . Hence there is an equilibrium position, between the flow center and the walls, where the particle is transported force-free. The equilibrium position is closer to the channel walls when the flow inertia increases as it was demonstrated theoretically in channel flow, first by \citep{schonberg1989inertial} up to $Re=O(100)$ and later by \cite{asmolov1999inertial} up to $Re=O(1000)$, assuming point-like particles.\\

The validation tests shown here were realized in square channel flow. Periodic boundary conditions were used in the flow direction ($Z$) and no slip at the walls (in $X$ and $Y$ directions). The ratio of the particle diameter to channel height was $d_p/H = 0.06$ and $0.11$. The channel length in the streamwise direction was equal to $28.8d_p$, where $d_p$ is the particle diameter. The grid distribution was set to ensure 8 grid points per particle diameter. The fluid flow was initially set to the steady solution of square channel flow, and a constant pressure gradient was applied in the $z$ direction. The particle was seeded at different $Y$ locations in the midplane ($X=H/2$). \\

\subsection{Particle freely moving in square channel flow}

\begin{figure} [h]
\includegraphics[width=0.45\linewidth]{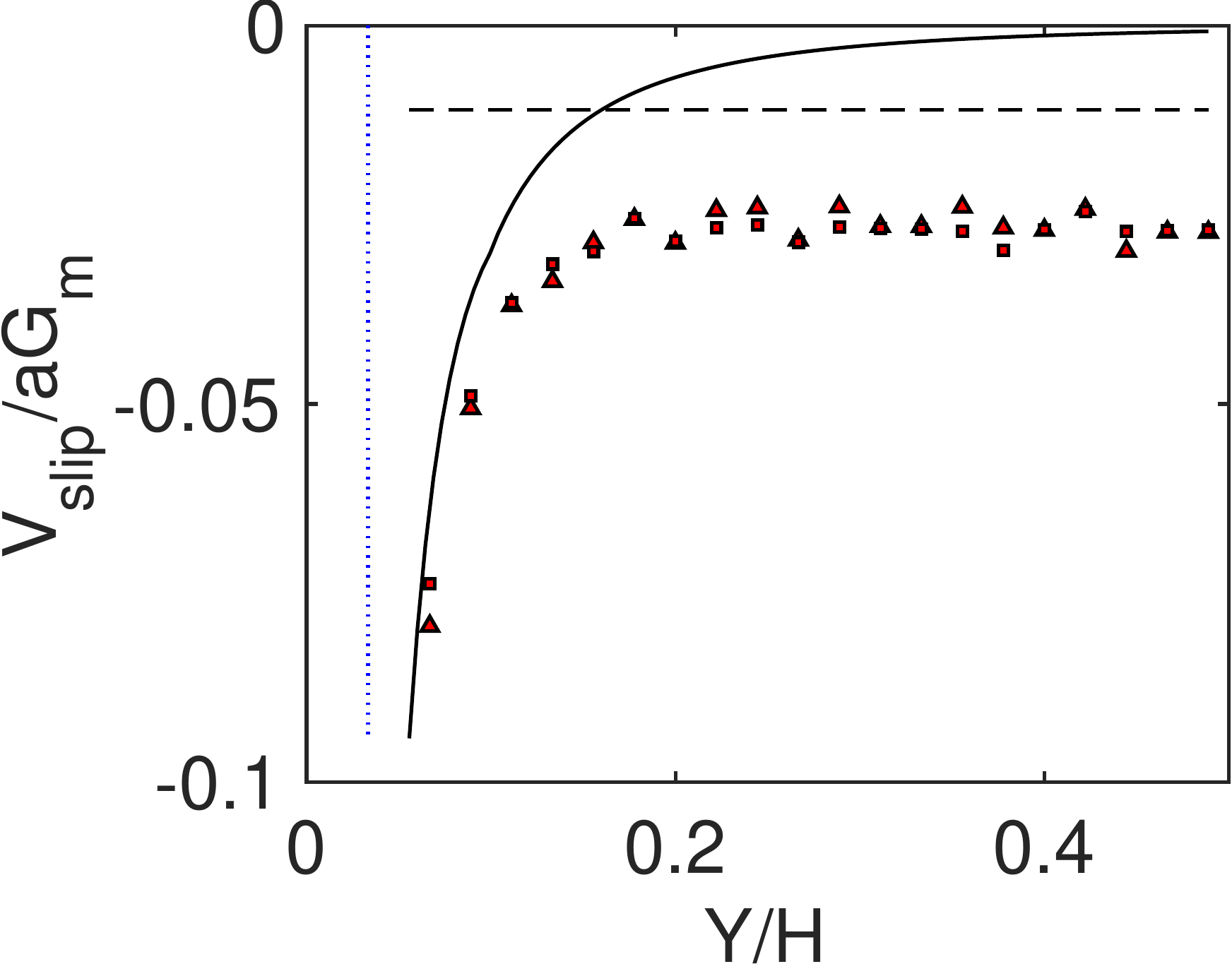}
\includegraphics[width=0.45\linewidth]{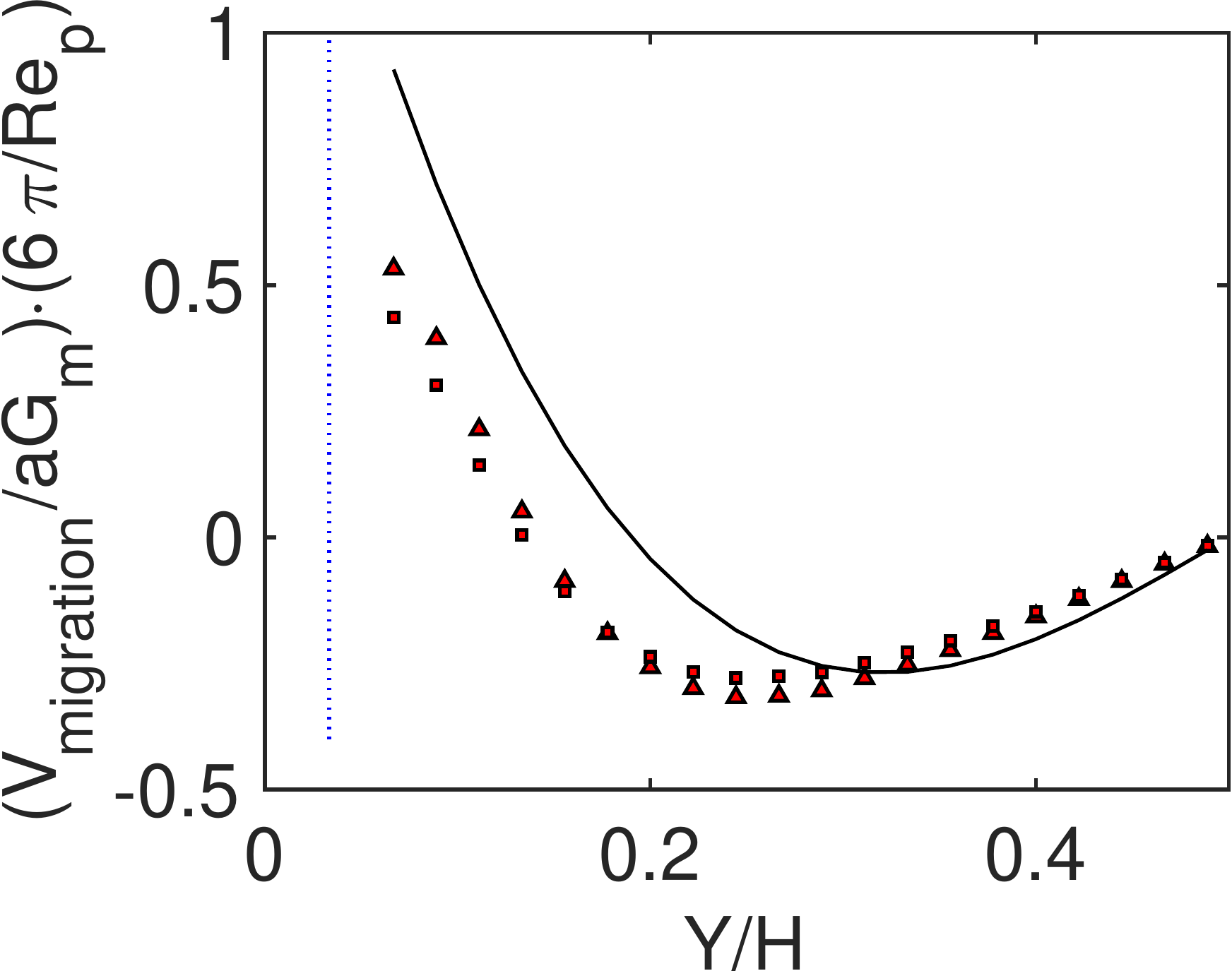}
\caption{Slip velocity (left panel) and migration velocity (right panel) vs particle position $Y/H$ for $Re = 13$ (triangles) and $Re = 39$ (squares) with particle size $d_p/H = 0.06$. The solid black curve shows the law proposed by ~\cite{asmolov2018inertial}. The vertical dotted line indicates a distance from the wall equal to the particle radius. The horizontal dashed line in left panel represents the Faxen Correction. }\label{fig:vmigration}
\end{figure}

In the first test, the particle was moving freely during approximately $10 a^2/\nu$, before its streamwise and wall normal velocities were recorded ($a$ is the particle radius).The streamwise slip and wall-normal particle velocities are shown in fig.~\ref{fig:vmigration}. The two velocity components are compared, at channel Reynolds number $Re = 13$ and $39$, to theoretical expressions for a point-like particle in 2D Poiseuille flow (see the summary on this in \citet{asmolov2018inertial}). For the smallest particle size, the effect of the flow three-dimensionality on the particle motion is expected to be relatively small. The slip velocity is normalized by $aG_m$, where the shear rate $G_m = 4 U_m/H$ is calculated from the maximum velocity in the channel center $U_m$. The particle slip is not impacted by the flow inertia in this range of Reynolds numbers. The agreement with the theoretical velocity as derived from \citet{goldman1967slow} is acceptable near the wall. However, the slip does not vanish in the channel center because of the flow curvature, (this effect on the particle motion was written formally in Fax\'en laws at low Reynolds numbers). The slip magnitude is roughly twice of the Fax\'en correction $4 U_m a^2/(3H^2)$. The same observations were reported in the studies of \citet{loisel2015inertia} and \citet{asmolov2018inertial}, realized in 2D Poiseuille flow, using numerical simulations based on FCM and Lattice Boltzmann respectively. As for the migration velocity (scaled by $aG_m$ and $Re_p/6\pi$), its trend is similar to the prediction based on a point-like particle at low but finite Reynolds number \citep{vasseur1976lateral}. The shift to the left of the numerical points is a joint consequence of the flow being 3D instead of 2D and to the under-estimation of the hydrodynamic interaction between the particle and the wall. We note that this under-estimation seems to be only effective on the wall-normal direction and not on the slip parallel to the wall. \\

\subsection{Lift force computation}

The Force Coupling formulation allows solving the mobility problem, i.e., the particle is displaced and rotated under a given forcing. A neutrally buoyant particle is not subject to any external forcing. The direct calculation of the force pushing the particle to move across the flow streamlines is not possible, because unlike other particle-resolved methods, the Force Coupling Method does not guarantee the no-slip boundary condition on the particle surface, and therefore the surface traction cannot be directly calculated. Instead, an iterative algorithm is set to compute the wall-normal force that should be applied on a particle placed in a shear flow, in order to prohibit the particle motion in the wall-normal direction. \\

After recording the velocities of the freely moving particle in the previous test, a force was then applied on the particle, only in the wall-normal direction, in a way that ensures zero migration velocity. This force was applied to the particle motion through $\bm{F}^n _{ext}$ in \equeqref{eq:FCM4}. Its initial value was set to zero. The force was then updated at the iteration $k$ from the value at iteration $k-1$, according to a penalty method: $F_{ext}(k) = F_{ext}(k-1) - \lambda[6\pi\mu aV(k-1)]$. The iterations were stopped when $V(k)$ became very close to zero. $\lambda$ is an arbitrary constant, which should be chosen not very low in order to reduce the time needed for convergence and not very high in order to avoid numerical instability. Note that the computation of this force was first realized in the case of a particle placed near a wall in a linear flow. The details of this test are written in the appendix A. The similarity between this force applied to prohibit the particle wall-normal motion and the theoretical predictions of the quasi-steady lift force, led us to call it ``lift force" in this paper. \\

\begin{figure} [h]
\includegraphics[width=0.90\linewidth]{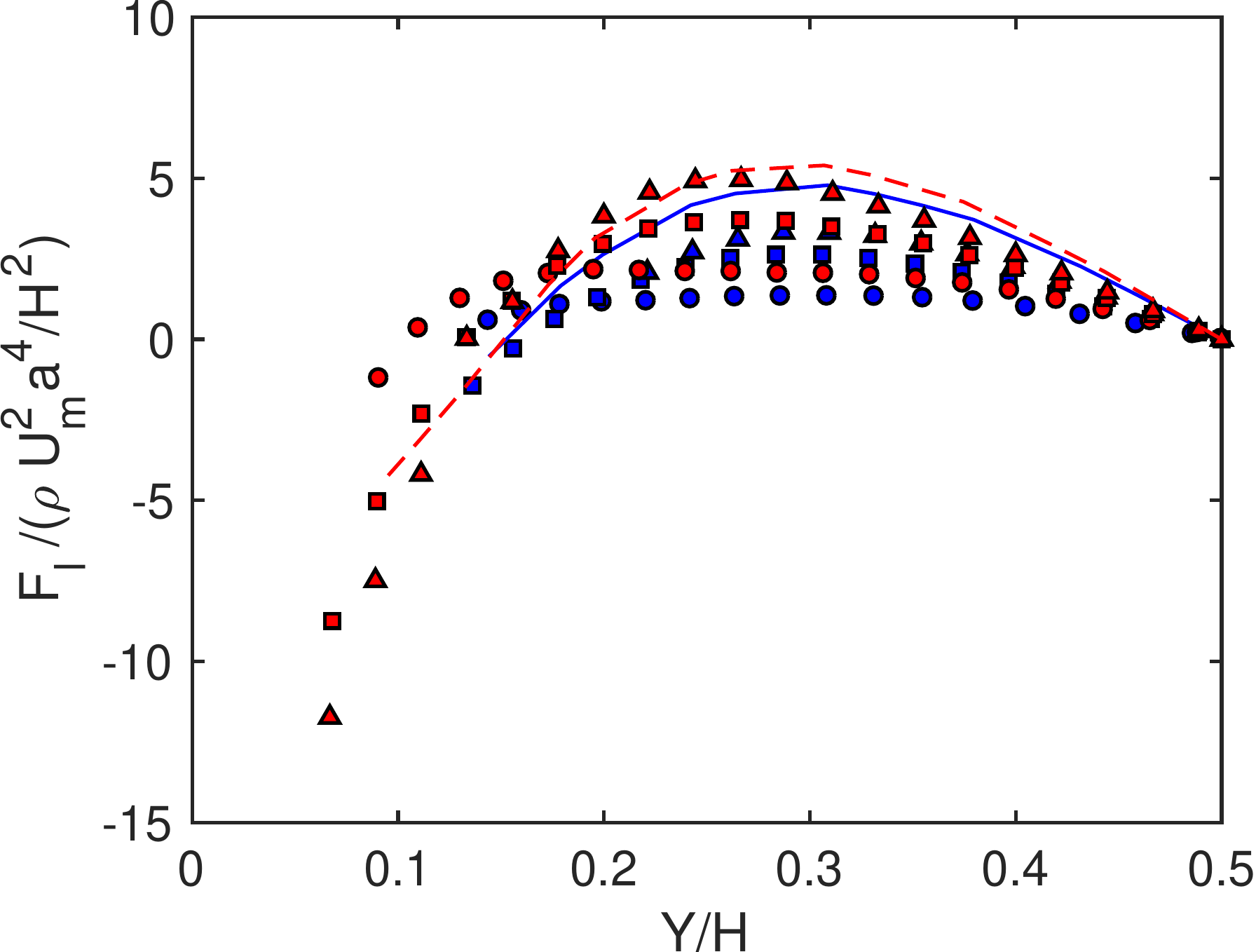}
\caption{Lift force scaled by $\rho U_m^2 a^4/H^2$, acting on a particle in a square channel flow versus the position of the particle in the y-direction (at $x = H/2$) for different ${Re}$. The red and blue symbols are the lift force from our simulation for particle diameter $d_p/H = 0.06$ and $0.11$, respectively. The symbols are for $Re = 120$ (circles), $38$ (squares) and $13$ (triangles). The corresponding $Re_p$ is between $0.05$ and $1.5$.  The red dashed and  blue solid lines are obtained from the solution of ~\cite{hood2015inertial} for particle diameter $d_p/H = 0.06$ and $0.11$, respectively.}\label{fig:lift_channel}
\end{figure}

Figure ~\ref{fig:lift_channel} shows the evolution of the lift force calculated $F_l/(\rho U_m^2a^4/H^2)$ as a function of the particle position, in the midplane ($X=H/2$) of the square channel flow. The particle radius $a$ is used in the force scaling. The negative sign indicates a force pushing the particle away from the wall. The numerical results, at different particle diameters and channel Reynolds numbers, are compared to the theoretical work of \citet{hood2015inertial}. Their work was developped in square channel flow geometry, assuming that the wall falls in the inner layer perturbed by the particle (weak inertial stresses compared to viscous stresses). The inertial lift force in the $x$ and $y$ directions was shown to depend on the particle radius in the form $F_l/(\rho U_m^2a^4/H^2) = C_{4}+C_{5}a/H$ where $C_{4}$ and $C_{5}$ are constants that depend only on the location of the particle. For the lowest Reynolds number ($Re=13$) and smallest particle size, the numerical force obtained by the FCM is in good agreement with the profile established in ~\cite{hood2015inertial}. Scaling the force by $\rho U_m^2 a^4/H^2$ is consistent in the channel center, but not near the channel wall. Still at $Re=13$, the dimensionless lift is lower when larger particle size is used. Note that, when the flow conditions are unchanged, the larger the particle the stronger is the impact of truncating higher order terms in the multipole expansion (quadrupole, sextupole...). When the Reynolds number is increased, the force calculated by the FCM deviates with respect to this scaling (it becomes lower), in a way coherent with the theoretical analysis based on matched asymptotic expansions \citep{schonberg1989inertial,asmolov1999inertial}. \\ 

%%%%%%% Microchannel Lift %%%%%%%%%%%

%%%%%%%%%%%%%Pair dynamics %%%%%%%%%%%%%%%%%%%%%%%%%%%%%%%%%%%%%%%

\section{4. Particle alignment in channel flow}

Here and in the following sections, the particles are freely transported by a square channel flow, unless otherwise stated. We considered particularly the square channel flow configuration to discuss the stability of particle alignment, because in the conditions of this paper, the equilibrium positions are stable and well established (at the midplane of the four channel walls) and trains have been well characterized by the experiments of Gao et al.~\cite{gao2017self}. The trajectory of a single particle migrating toward an equilibrium spot (as sketched in figure \ref{fig:sketch}), using the same numerical method, can be found in \cite{abbas2014migration}. When particles are randomly seeded in the simulation domain, they experience first a lateral motion i.e., perpendicular to the velocity iso-contours, then cross-lateral migration, i.e. parallel to the closest wall, till they reach equilibrium positions ~\cite{abbas2014migration}. Both processes are slow, the former stage being faster than the following one. The establishment length scale of particle migration is quite large ($O(1000H)$ at $Re=O(100)$), and the lower the Reynolds number the larger the establishment length is. In order to focus on the streamwise ordering process, the lateral and cross-lateral migration stages are bypassed by initially placing the particles near their equilibrium spots (in the midplane $x=H/2$). During the simulations, the particles were observed to remain in this symmetry plane. \\

The operating conditions consist of a channel Reynolds number $O(100)$, a particle confinement $\DP/H$ in the range $0.077-0.14$, and a solid volume fraction less than $1\%$. The flow was resolved using a uniform mesh grid with $78 \times 78$ grid points in the square cross-section, to ensure 8 grid points per particle diameter for a reasonable numerical accuracy. We carefully verified that the box length $L$ does not impact the results shown here ($29 \leq L/\DP \leq 60$, $\DP$ being the particle diameter). \\

The streamlines around a single particle are first shown, since they contribute to the alignment process. Then particle relative trajectories are used in order to show stable assemblies when small number of particles is involved (two, or three) and unstable assemblies as soon as the number of particles becomes "large".

\subsection{Streamlines around a single particle at equilibrium}

\begin{figure} [h!]
\hspace{-0.5cm}
\vspace{-0.8cm}
\begin{center}
{\includegraphics[trim=0 0 60 24, clip, width=0.35\linewidth]{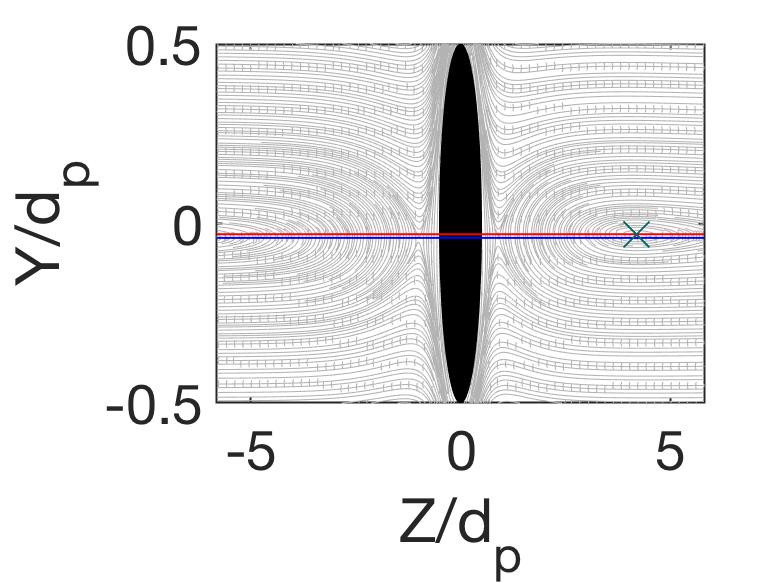}}
\put( 5,3){\color{red}{\vector(1,0){023}}}
{\includegraphics[trim=90 0 60 24, clip, width=0.31\linewidth]{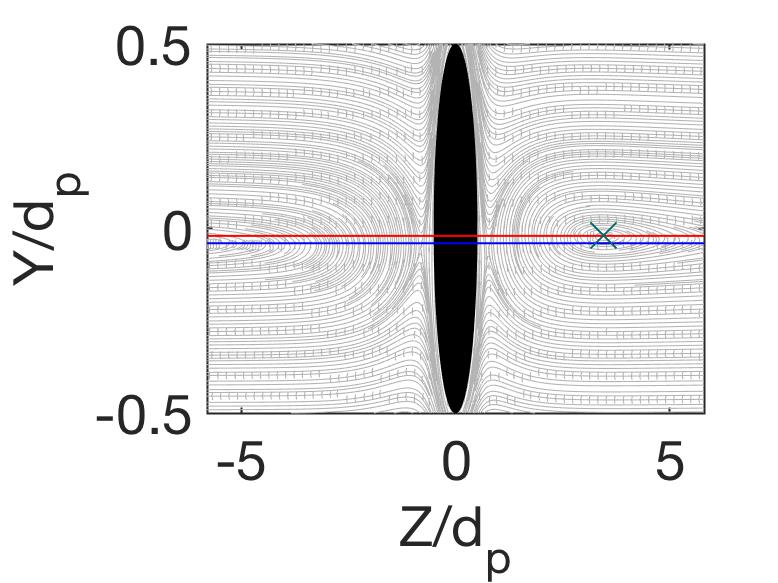}}
{\includegraphics[trim=90 0 60 24, clip, width=0.31\linewidth]{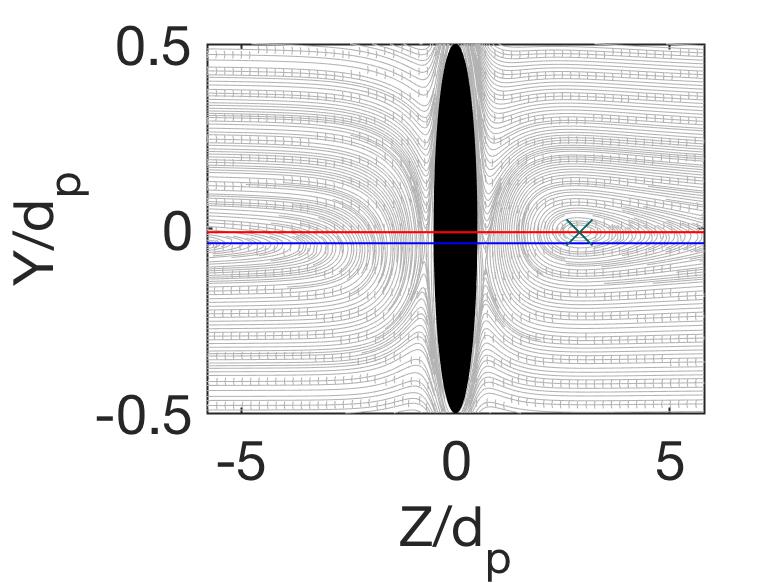}}
\begin{picture}(0,0)
\put( -183,56){\color{red}{\vector(1,0){020}}}
\put( -107,56){\color{red}{\vector(1,0){020}}}
\put( -027,56){\color{red}{\vector(1,0){020}}}
\end{picture}
\end{center}
\vspace{-0.6cm}
\caption{Zoom on the flow streamlines in the ($YZ$) frame of a particle at equilibrium in a square channel flow. The black region illustrates the particle (stretched for convenience). $\ReP = 0.5$ (left), $1.5$ (center) and  $3.0$ (right). The two stagnation points (indicated with crosses) get closer to the particle surface as $\ReP$ increases. The fore-aft asymmetry increases with $\ReP$. Red arrows show the flow direction.}\label{fig:zero_single}
\end{figure}
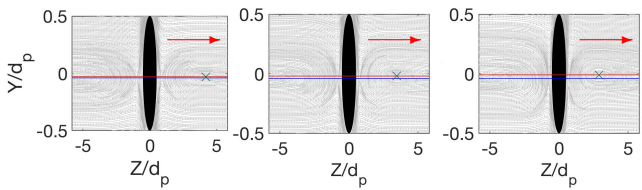

\Fig{fig:zero_single} shows the flow streamlines in the ($YZ$) frame attached to a single particle located at an equilibrium spot, for different Reynolds numbers at particle confinement $d_p/H=0.11$ (the images are stretched and zoomed for format convenience). The saddle points both in front of and behind the particle take place in the presence of shear as soon as the flow inertia is finite at the particle scale \cite{Subramanian2006}. Whereas the reverse streamlines induced by a particle in a channel flow are open under Stokes flow condition, they are of spiralling nature at finite inertia ~\cite{lee2010dynamic}. The stable spirals act as attractor regions. The centers of these forward and backward spirals are closer to the particle surface as $\ReP$ increases and the size of this attractive region becomes wider. The two horizontal lines added to these figures show that the gap (in the $y$-direction) between the forward and backward attractors also increases with $\ReP$, in relation with the symmetry breakup at finite flow inertia. \\

%%%%%%%%%%%%%%%%%
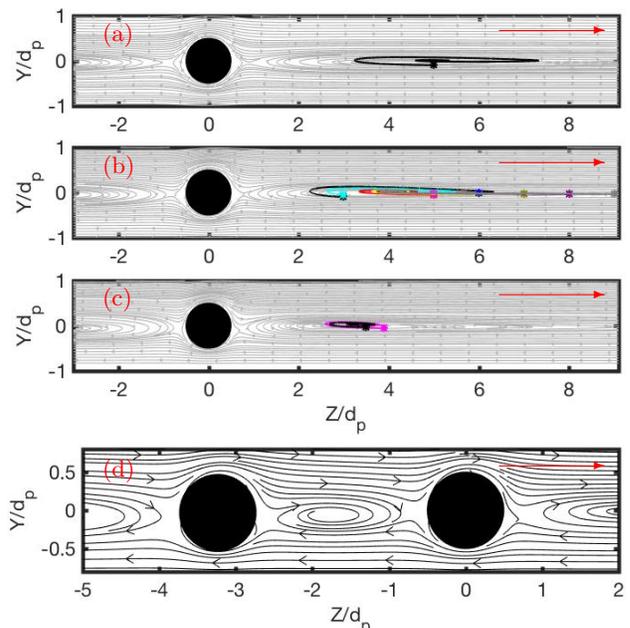
\begin{figure}
\vspace{-0.0cm}
\begin{center}
{\includegraphics[trim=27 220 30 220, clip, width=1.00\linewidth]{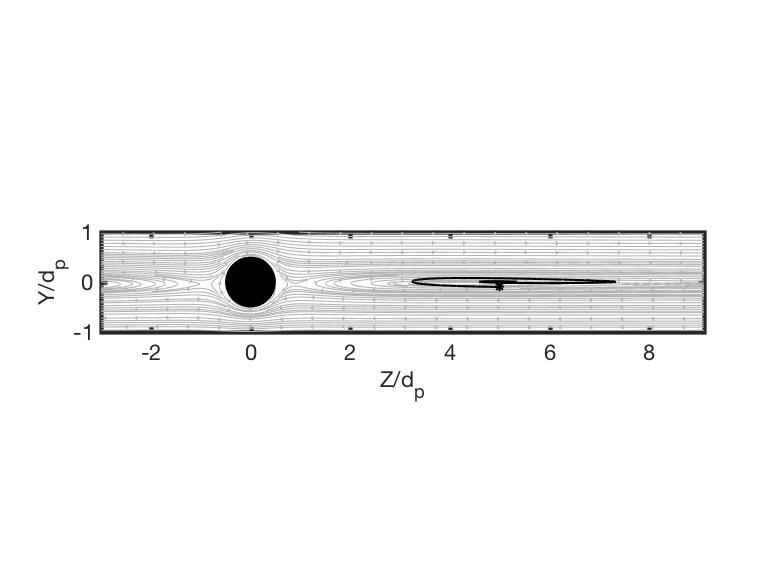}}
\put( -210,035){\color{red}{(a)}}
\put( -060,039){\color{red}{\vector(1,0){040}}}\\
{\includegraphics[trim=27 220 30 220, clip, width=1.00\linewidth]{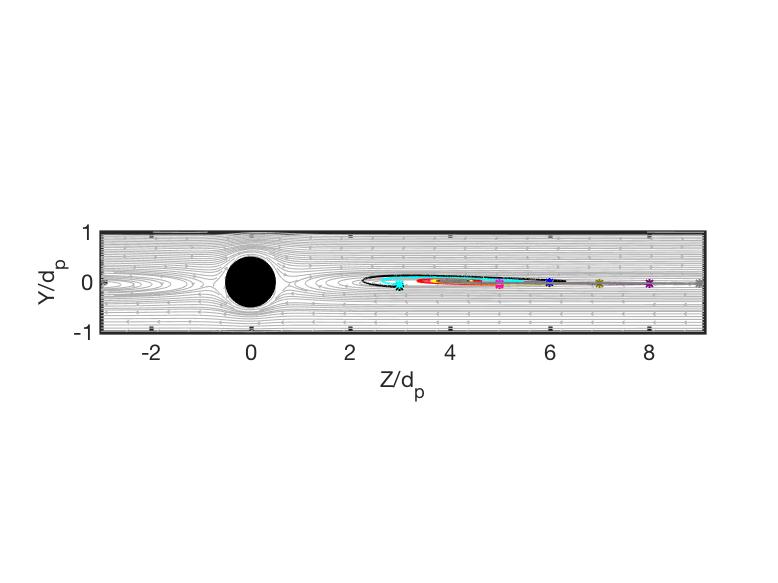}}
\put( -210,035){\color{red}{(b)}}
\put( -060,039){\color{red}{\vector(1,0){040}}}\\
{\includegraphics[trim=27 180 30 220, clip, width=1.00\linewidth]{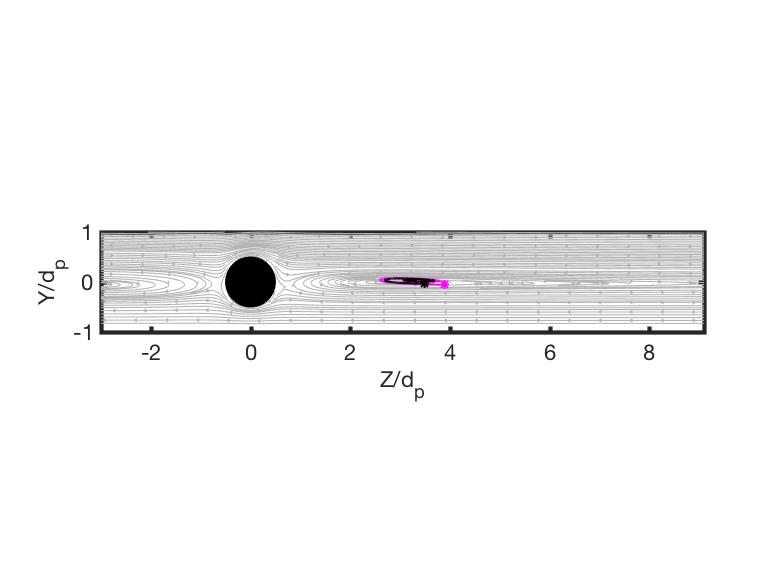}}
\put( -210,049){\color{red}{(c)}}
\put( -060,053){\color{red}{\vector(1,0){040}}}\\
{\includegraphics[trim=15 160 30 200, clip, width=1.00\linewidth]{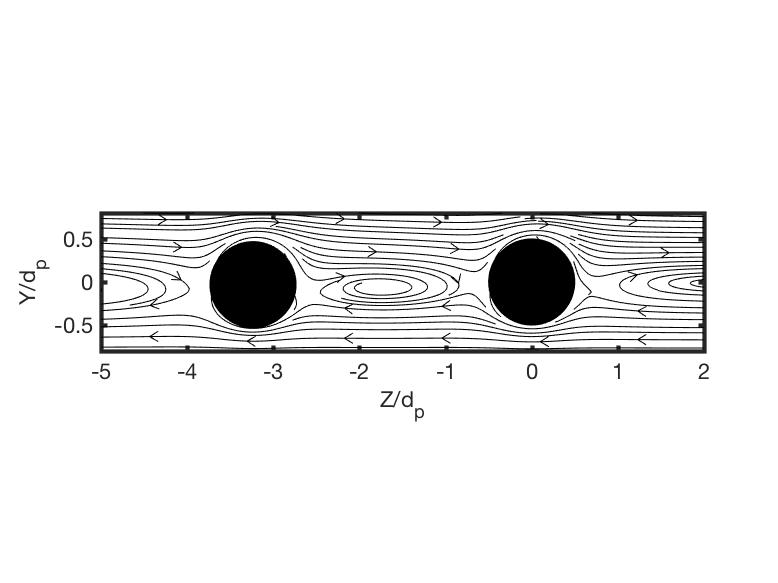}}
\put( -210,060){\color{red}{(d)}}
\put( -060,064){\color{red}{\vector(1,0){040}}}
\end{center}
\vspace{-0.8cm}
\caption{a, b and c) contain the pair-dynamics in a two-particle train for $\ReP = 0.5$, $1.5$ and  $3.0$ respectively. The trajectories of the leading particle with respect to the lagging one are overlaid on the streamlines around a single particle. The initial position of the center of the leading particle is shown with asterisks. Red arrows show the flow direction. (d) Flow streamlines (in the frame of leading particle) around a stable pair of particles obtained for $\ReP=3$ , showing a stable small recirculation connecting both particles.}\label{fig:Pair_dyn}
\end{figure}

\subsection{Stable particle assembly}

When two particles are found close to each other near an equilibrium spot (as illustrated in \Fig{fig:sketch}), they are located in-plane, i.e. in the plane parallel to the flow and to one wall-normal direction. Studies dedicated to the interaction of a particle pair in a linear flow at finite flow inertia \citep{yan2007hydrodynamic,Haddadi2015}, show that the relative trajectory of the lagging particle with respect to the leading one is open (resp. reversed) when the shift $\delta Y$ in the position of the particle centers in the wall-normal direction is large (resp. small). \\

\Fig{fig:Pair_dyn} shows the relative trajectory of a leading particle (particle 2) with respect to the lagging one (particle 1) in the square channel flow at $d_p/H=0.11$. The trajectory of particle 2 is trapped in a basin of attraction, nearby the forward attractor of particle 1, following a spiralling motion. At $\ReP=0.5$, the trajectory of particle 2 with respect to particle 1 is similar to the streamlines around the freely rotating particle 1 (when isolated). The first part of the relative trajectory (in fig. \ref{fig:Pair_dyn}) is of reversing nature since $\delta Y=Y_{P2}-Y_{P1}$ is initially small and negative. However, after reversing, particle 2 does not go off to infinity. The inertia-induced lift induces cross-streamline relaive motion, which is coupled to the trajectory reversal leading to an equilibrium spacing between the particles. This type of interaction of two finite $Re_p$ spheres, is an additional aspect compared to the open and reversed trajectories in linear flow. It involves the quadratic nature of the flow and the proximity of the particles to the wall. When $\ReP$ increases, particle 2 converges faster toward equilibrium, the convergence pathway depending on the initial position of particle 2 with respect to particle 1. \subfig{fig:Pair_dyn}{b} ($\ReP=1.5$) shows that if the relative position is chosen carefully, the leading particle converges toward the basin of attraction even if the initial distance is as far as $9\DP$. At equilibrium, the vortex in front of the lagging particle and the one behind the leading particle connect to form one close vortex as shown is \subfig{fig:Pair_dyn}{d}. \\

%%%%%%%%%%%%% Asymmetry: Three particle train%%%%%%%%%%%%%%%%%%%%%%%%%%%%%%%%%%%%%%%
%%

\begin{figure}[h]
\hspace{-0.5cm}
\begin{center}
\includegraphics[trim=50 150 100 240, clip, width=1.00\linewidth]{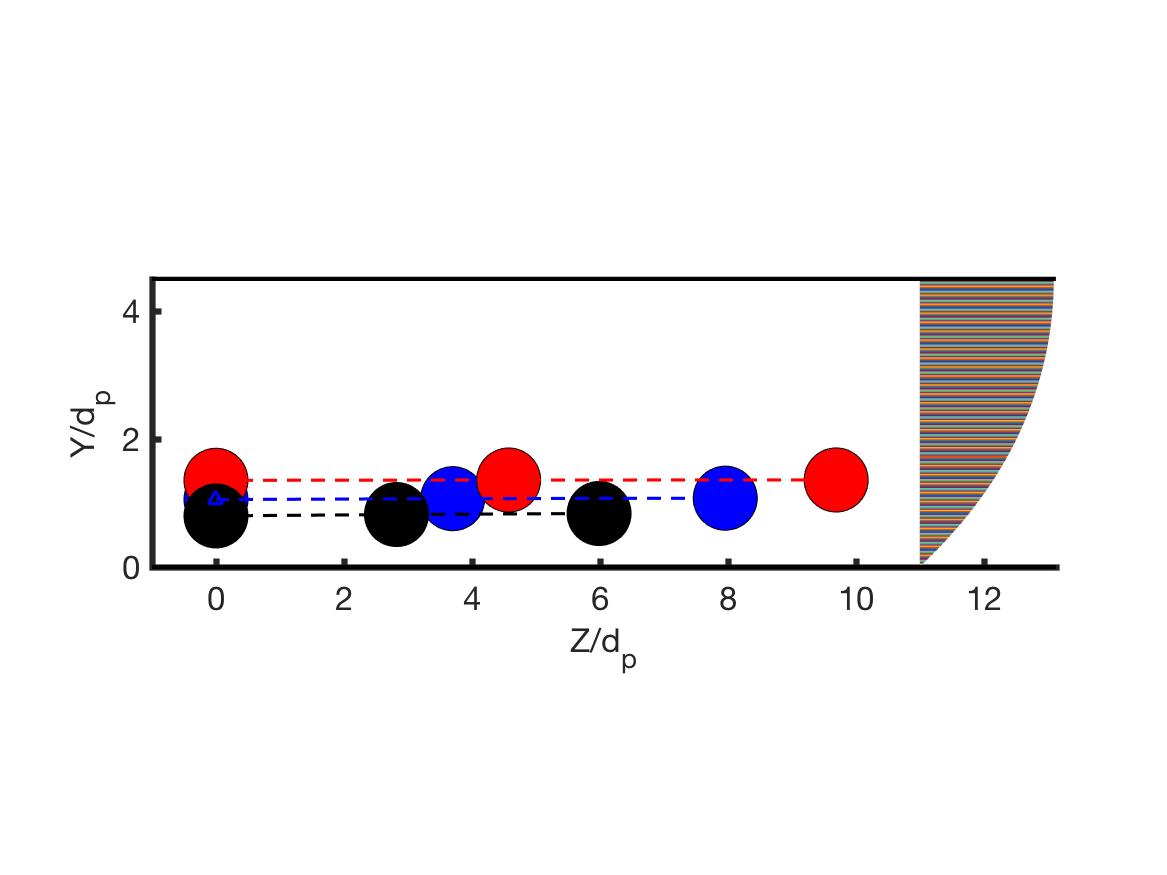}
\put( -130,094){\vector(1,0){085}}
\put( -115,098){Flow direction}
\put( -202,094){\bf (a)}\\
{\includegraphics[width=0.8\linewidth]{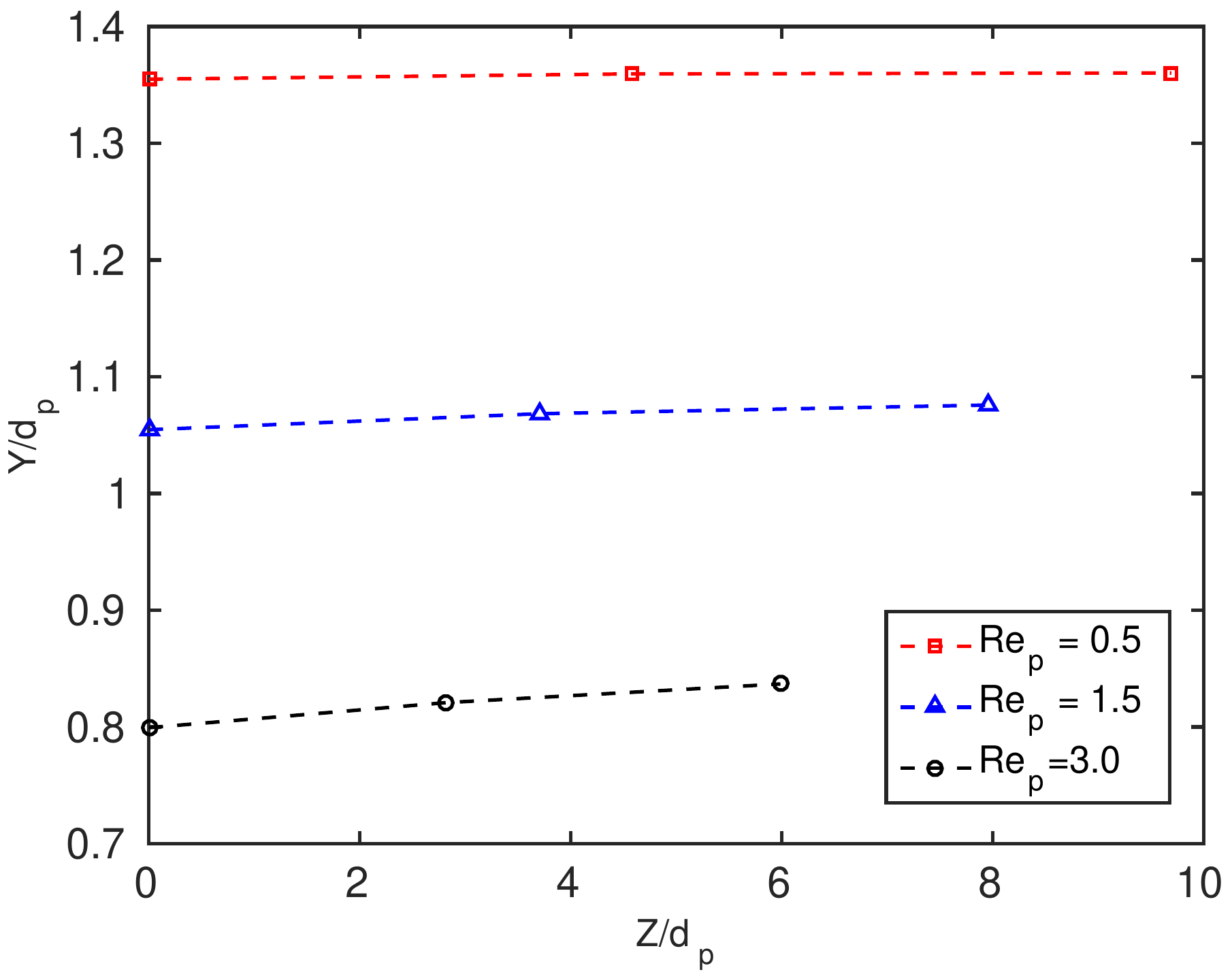}}
\put( -160,127){\bf (b)}\\
\end{center}
\vspace{-0.6cm}
\caption{ (a) Steady configuration of three-particle trains at different Reynolds numbers in the ($YZ$) plane. Particles in the same train have the same color and are linked with lines for eye guidance. $\ReP = 0.5$ (red), $1.5$ (blue) and $3$ (black). The z-coordinate of the lagging particle in a train is set arbitrarily to 0. The velocity profile is represented on the right of the figure. (b) Plots representing the center positions of the three-particle trains at different Reynolds numbers (same colors as in (a)). }\label{fig:three_part}
\end{figure}

We realized the same type of simulations with three particles. \subfig{fig:three_part}{a} shows the particle trains at equilibrium for three different $\ReP$ at $d_p/H=0.11$. The streamwise position of the lagging particle is set arbitrarily to zero for this figure. The inter-particle spacing and average train distance from the wall both decrease when $\ReP$ is increased. These trains are not perfectly aligned in the flow direction, but they are relatively inclined as shown in \subfig{fig:three_part}{b}. This has been also observed experimentally by Matas et. al.~\cite{matas2004trains} at higher $\ReP$ in a tube flow, while the inclination is absent at smaller $\ReP$. The evolution of the train inclination with the Reynolds number is coherent with the gap between forward and backward stagnation points shown in \Fig{fig:zero_single}, which increases with $\ReP$. In addition, it can be noted that the spacing between the leading and second particle is $10\%$ greater than between the second and third (lagging) one in all cases.\\

\begin{figure}[h!]
\hspace{-0.5cm}
\begin{center}
\includegraphics[width=0.8\linewidth]{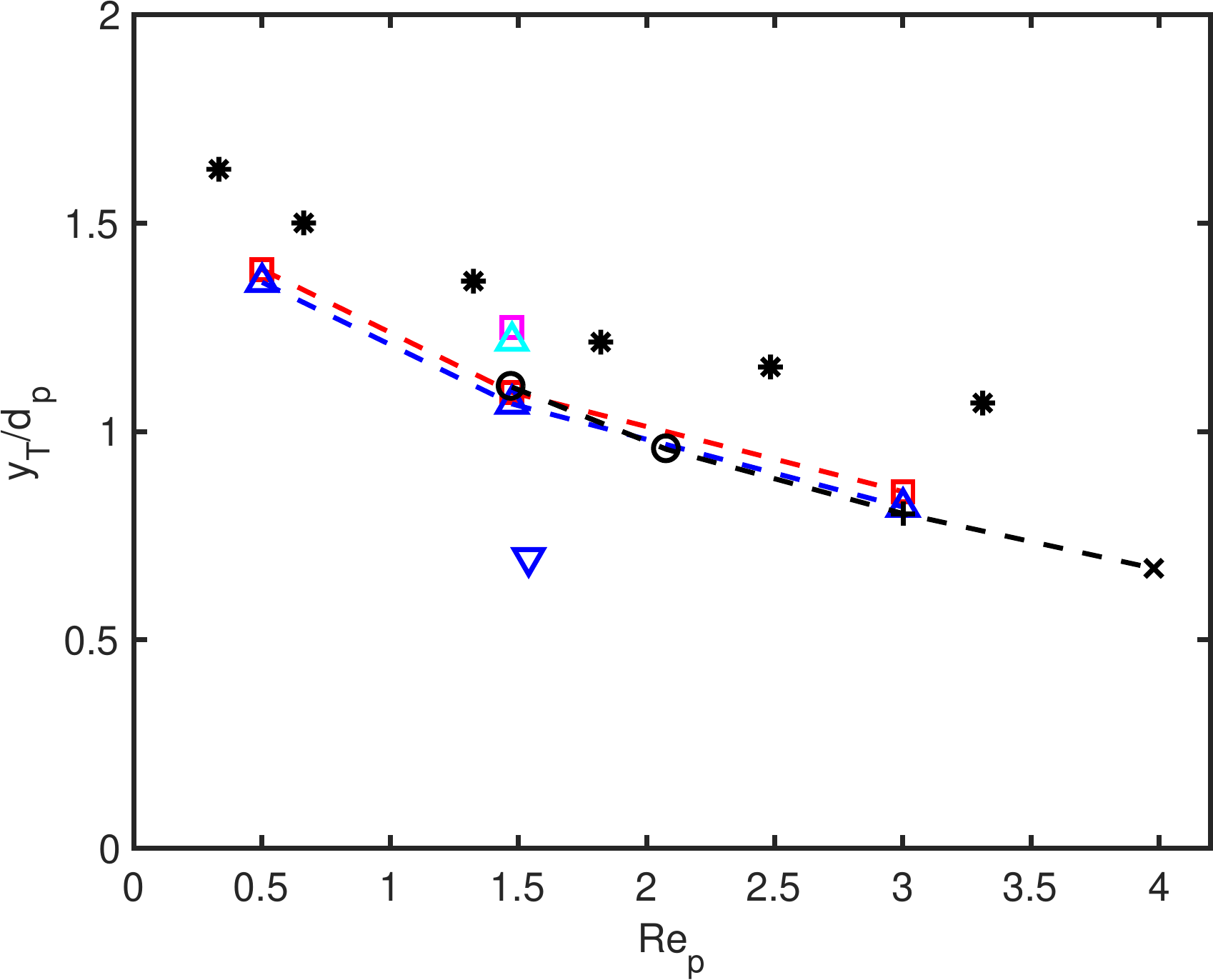}
\put( -060,127){\bf (a)}\\
{\includegraphics[width=0.8\linewidth]{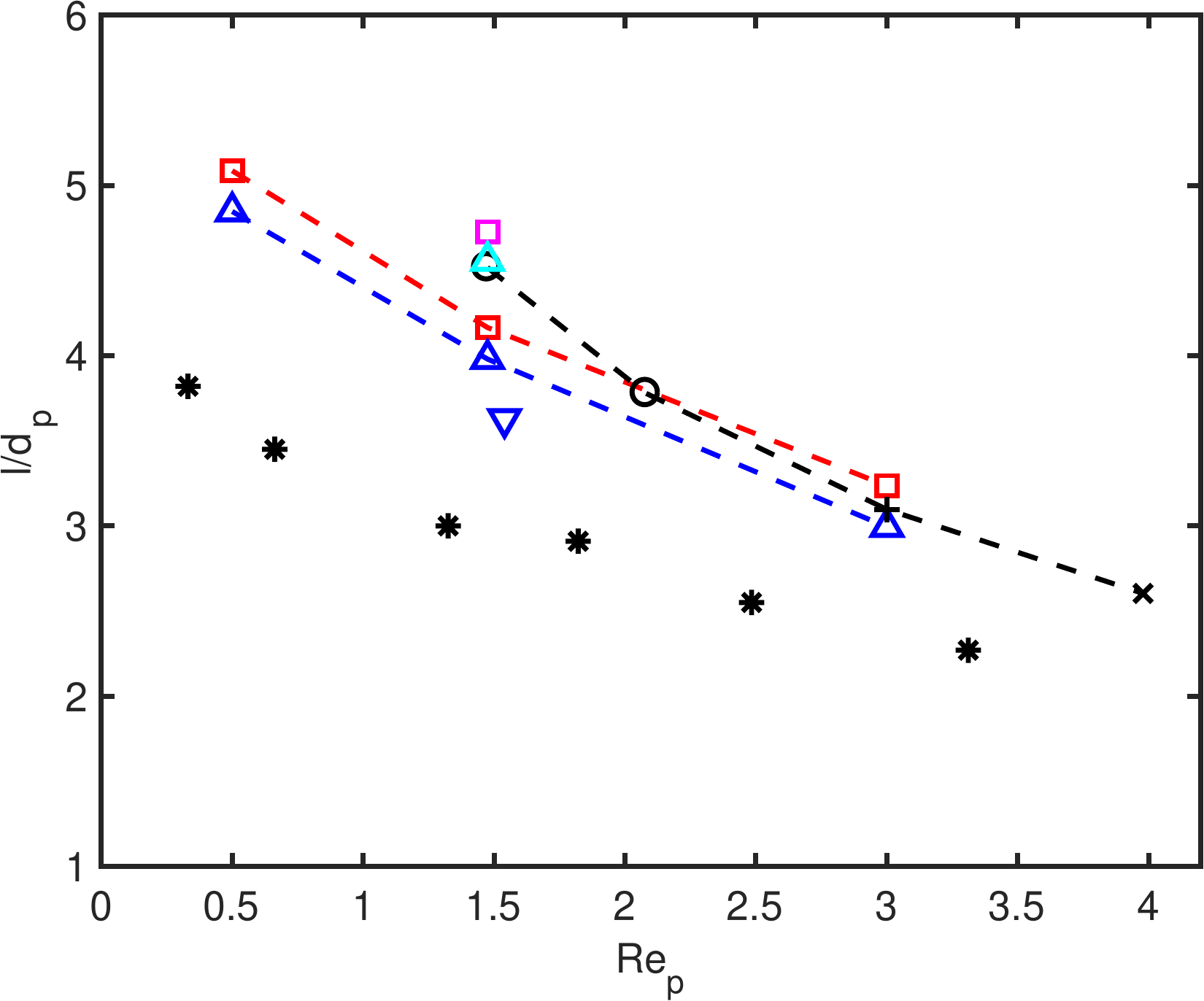}}
\put( -060,127){\bf (b)}\\
\end{center}
\vspace{-0.6cm}
\caption{Properties of stable particle trains. (a) shows the distance between the train barycentre and the closest wall $y_T$. (b) shows the average interparticle distance $l$. Red Squares, upward-pointing triangles and black plus symbols are for 2, 3 and 4 particle assembly in square channel with $\frac{d_p}{H}=0.11$. Blue downward-pointing triangles and black circles are for 3 particle and 4 particle assembly in square channels with $\frac{d_p}{H}=0.14$ and $0.08$ respectively. Black cross is for 4 particle in rectangular channel ($\frac{d_p}{W}=0.09$). Black stars are from the experiments of \cite{gao2017self} realized in square channel flow with $\frac{d_p}{H}=0.11$. The magenta square and cyan triangle are obtained for 2 and 3 particle trains, using a twice finer numerical resolution in a square channel with $d_P/H = 0.11$}.\label{fig:train_stat}
\end{figure}

The distance between the train barycentre and the closest wall $y_T$, as well as the average interparticle distance at equilibrium $l$ are plotted in \Fig{fig:train_stat}. Most of the simulations were realized with square channel flow and $d_p/H=0.11$. When $\ReP$ is increased, the train gets closer to the channel wall (similarly to the single particle) and the average inter-particle distance decreases. The decrease of the average distance with $\ReP$, observed similarly in the experiments ~\cite{matas2004trains,kahkeshani2016preferred,gao2017self}, is consistent with the fact that the basin of attraction is closer to the particle surface, when the particle Reynolds number is increased (fig. \ref{fig:zero_single}). \Fig{fig:train_stat} contains also information on particle assembly when the number of particles per train is increased. At a given $\ReP$, the train gets slightly closer to the wall when the number of particles per train increases. The average interparticle distance seems to slightly decrease as well. The train statistics are compared to the experimental ones of \citet{gao2017self} which were realized in similar conditions. The trend of the trains statistics with respect to $\ReP$ is similar. There is a uniform shift between experimental and numerical results. The discrepancy of train positions at equilibrium is almost suppressed when the mesh resolution is twice finer. However the shift persists for the average interparticle spacing. This issue deserves to be further delved into in the future if precise information on the stability of particle alignment is needed. It requires the computation of the interaction between several particles in channel flow, at identical operating conditions, using different available (at least numerical) methods.  \\

%%%%%%%%%%%%% Unstable : Four particle train%%%%%%%%%%%%%%%%%%%%%%%%%%%%%%%%%%%%%%%
%%
\subsection{Unstable assembly}

\begin{figure}
\vspace{-0.0cm}
\begin{center}
{\includegraphics[trim=27 180 30 220, clip, width=1.00\linewidth]{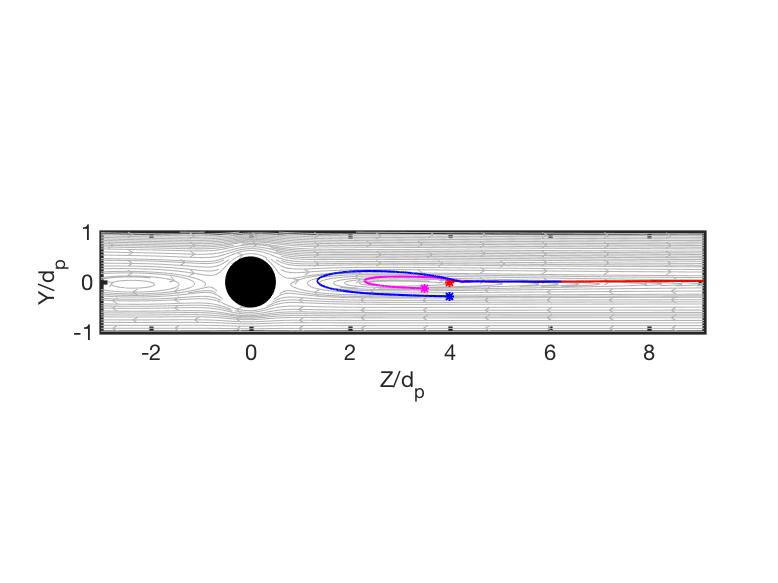}}
\put( -090,050){\color{red}{\vector(1,0){070}}}
\put( -081,052){\color{red}\scriptsize{Flow direction}}
\put( -202,048){\bf {\color{red}(a)}}\\
\end{center}
\includegraphics[width=0.80\linewidth]{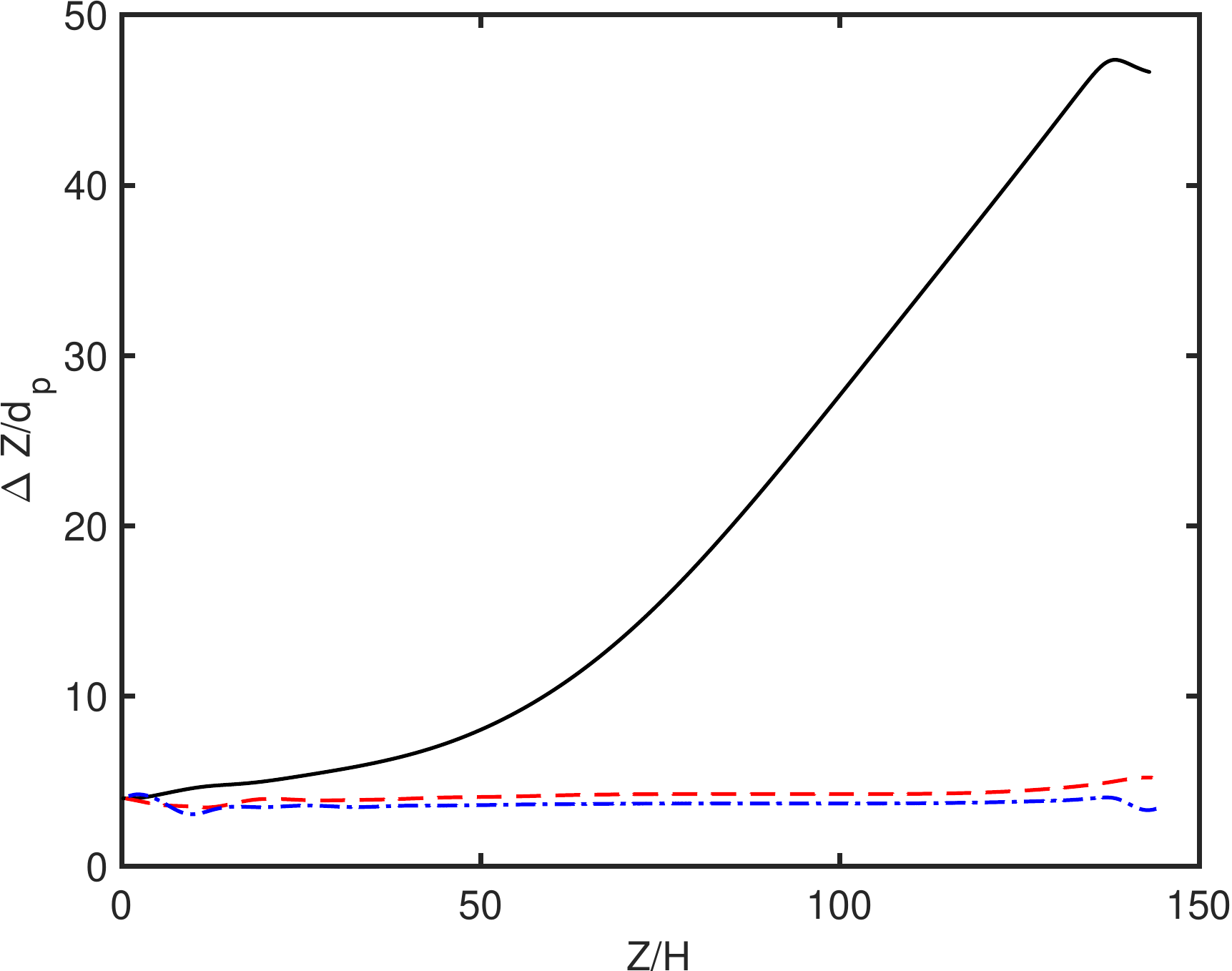}
\put( -160,127){\bf (b)}\\
\caption{(a) Trajectories of the fourth particle placed in front of a stable three-particle train at $\ReP=1.5$ and $d_p/H = 0.11$ for different initial positions. These trajectories are relative to the (new) leading particle of the remaining stable three-particle train (at $\ReP = 1.5$). The trajectories are overlaid on the streamlines, in ($YZ$) plane. (b) contains the evolution of the relative particle spacing in the train ($\Delta Z$).  Black, red and blue lines correspond to the distance between front-second, second-third and third-fourth. The $Z$ coordinate corresponds to the streamwise position of the front, second, and third particles for each curve respectively. }\label{fig:four_part}
\end{figure}

When a fourth particle is seeded close to the three-particle train of \Fig{fig:three_part} ($\ReP=1.5$), either in front of or behind the train, the front particle is lifted up, leaving a stable three-particle train behind it. The relative trajectory of a fourth particle placed in front of a three-particle stable train is shown in \subfig{fig:four_part}{a}. Several initial configurations led to the same result. Even if the leading particle tends to follow the reversed streamlines in a first stage, it does not converge toward the attractor. The departure of the front particle can be found in \subfig{fig:four_part}{b} from the evolution of the spacing between the front and second particles (in black lines). A video sequence of particle positions in the channel for this case is shown in the supplementary material~\cite{supmat}. The particle that leaves the train reaches an equilibrium position $y_P$ located slightly further from the wall than the position $y_T$ of the train left behind and is thus slightly faster. Since we used periodic boundary conditions in the flow direction, the particle that leaves the train from the front joins it from behind, the new leading particle leaves in turn, and so on. The same observations were noted for trains with the larger number of particles. \\

Note that this instability starts to be discernible after the particles travel a long distance downstream, (i.e. $\approx 30-40H$ in the case of figure \ref{fig:four_part}). It is probably for this reason that \cite{lee2010dynamic} have observed a shift of the distribution of inter-particle distances toward larger values in the measurements at different distances from the channel inlet, without observing any change in inter-particle spacings within images of dimensions of $O(10H)$. In order to detect the eventual departure of the leading particle from the train structure by optical techniques, it would be required to use either two synchronized cameras at different streamwise positions, or a camera frame following a long-lived train that contains a large number of particles.

%%

%%%%%%%%%%%%%%%%%%%%%%%%%%%%%%%%%%%%%%%%%%%%%%%%%
\section{5. Discussion on the destabilizing mechanism}

\begin{figure}
\hspace{-0.5cm}
\begin{center}
{\includegraphics[width=0.90\linewidth]{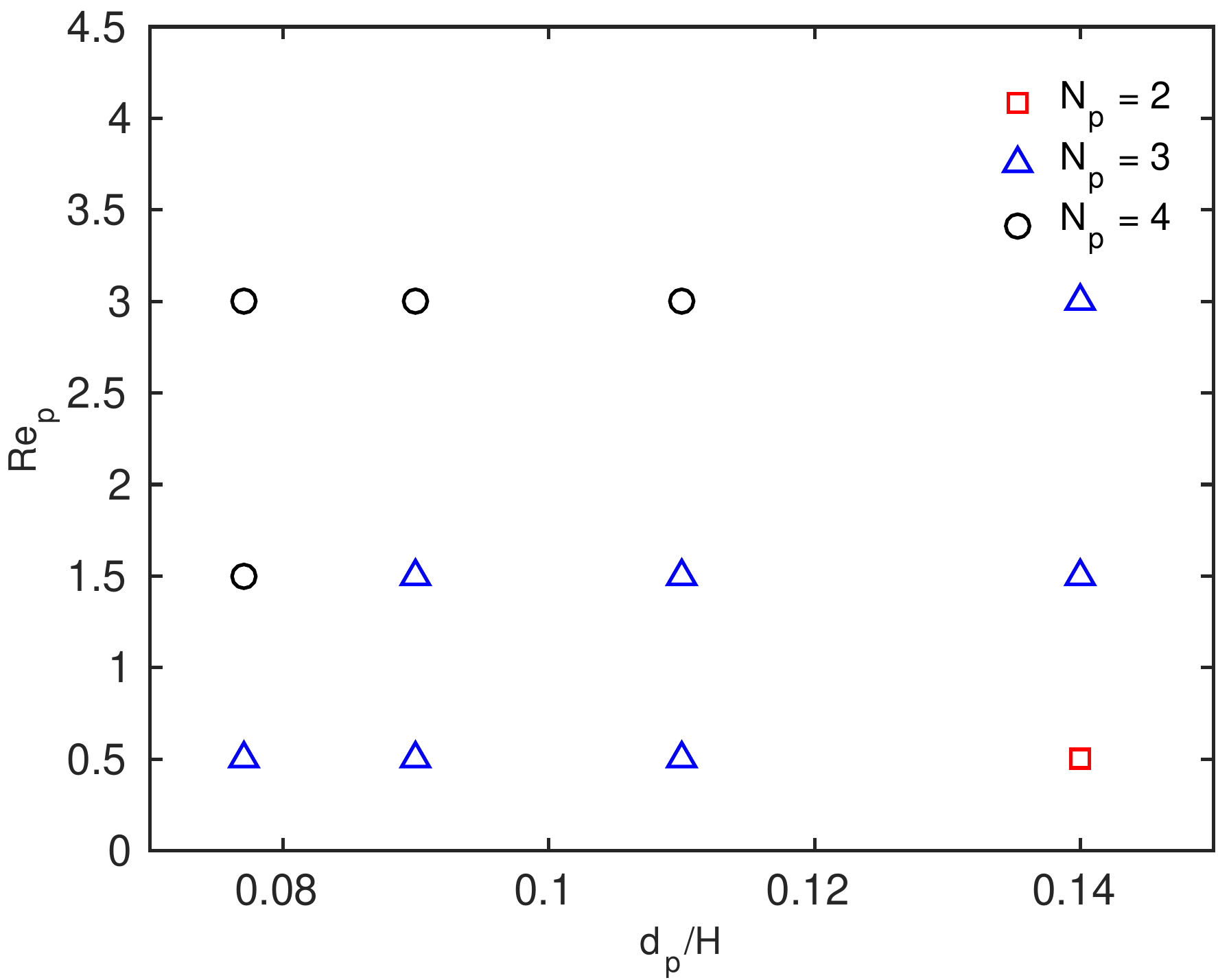}}
\end{center}
\caption{A diagram showing the maximum number of particles in a stable train for different particle Reynolds number $\ReP$ and confinement $\DP/H$ in a square channel flow, as predicted from the FCM simulations.  }\label{fig:phase}
\end{figure}
%%

%%%%%%%%%%%%%%% Phase %%%%%%%%%%%%%%%%%%%%%%%%%%%

The maximum number of stable particles in a train can be tuned by changing the particle confinement and/or the fluid inertia, as summarized in \Fig{fig:phase}. Note that in order to change $\DP/H$ while keeping constant the particle Reynolds number, the channel Reynolds number should be changed accordingly. It is clear from  \Fig{fig:phase} that both $\ReP$ and confinement play an important role. For rectangular channel cross-section, the confinement is defined as the ratio between the particle diameter and the channel height or width, whichever is smaller (setting the largest velocity gradients). The number of particles stable in a train increases when the flow inertia is increased and when the particle size is decreased. It is striking to note that the maximum length of stable train structures is approximately equal to the channel hydraulic diameter ($H$ for square channels).   \\

\begin{figure}
\hspace{-0.5cm}
\begin{center}
{\includegraphics[width=0.80\linewidth]{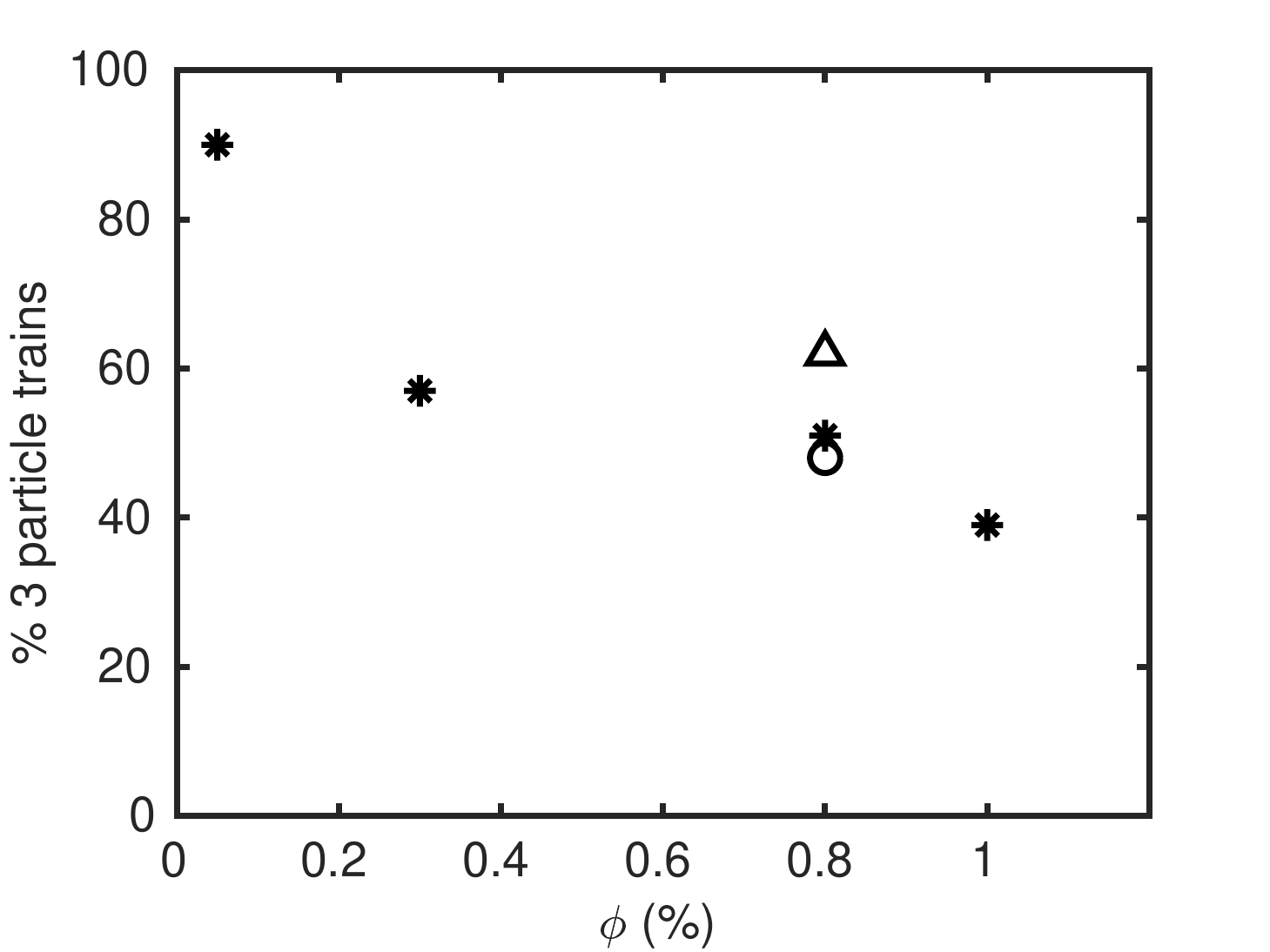}}
\end{center}
\caption{Probability of finding trains with 3 aligned particles, obtained from post-processing the experiments of \cite{gao2017self}, versus the suspension concentration. The probability is the highest at low concentration. Different symbols are for different $Re_p$: triangle, stars and circle are for $Re_p= 0.12$, 0.7 and 1.8. These results are for $d_p/H=0.11$.}\label{fig:perc_exp}
\end{figure}

The numerical results encouraged us to revisit the statistics of \citet{gao2017self} realized on experiments with particle size $d_p/H=0.11$ in square channel flow. The histograms of the number of aligned particles in a single train exhibit a very sharp peak at $N_p=3$, for $Re_p$ ranging from 0.1 to 3 and for particle concentration between 0.02 and $1\%$. The percentage of trains constituted of three particles is shown in figure \ref{fig:perc_exp} as a function of the suspension concentration (defined as the solid volumetric fraction). This figure shows that most of the trains are constituted of three aligned particles at low concentration ($\phi=0.08\%$). A longer train detected in the experiments (with imaging techniques \citep{gao2017self}) as such, might be the result of a transient alignment. As the concentration increases, the percentage of three-particle trains decreases but remains dominant. The concentration has the dual effect of both increasing the number of particles available for alignment, and the dispersive hydrodynamic interactions between them.  \\

\begin{figure*}
\begin{center}
{\includegraphics[trim=0  0 0 0, clip, width=0.35\linewidth]{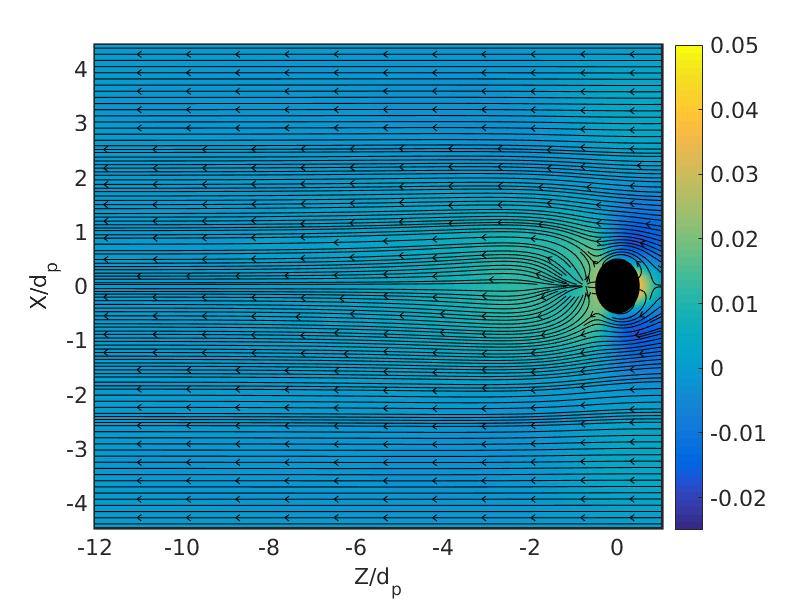}}
\put( -150,110) { \colorbox{white}{ \textcolor{black}{\bf (a)} } }
{\includegraphics[trim=60  0 0 0, clip, width=0.32\linewidth]{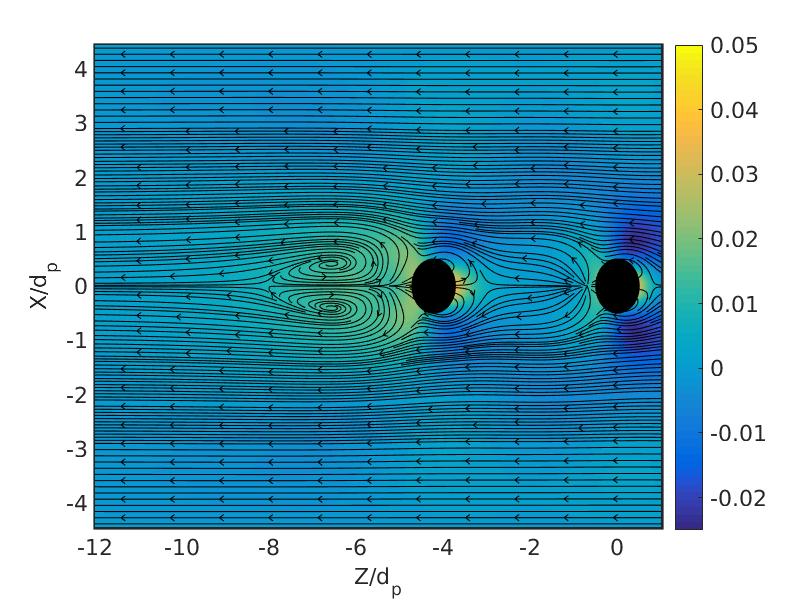}}
\put( -150,110) { \colorbox{white}{ \textcolor{black}{\bf (b)} } }
{\includegraphics[trim=60  0 0 0, clip, width=0.32\linewidth]{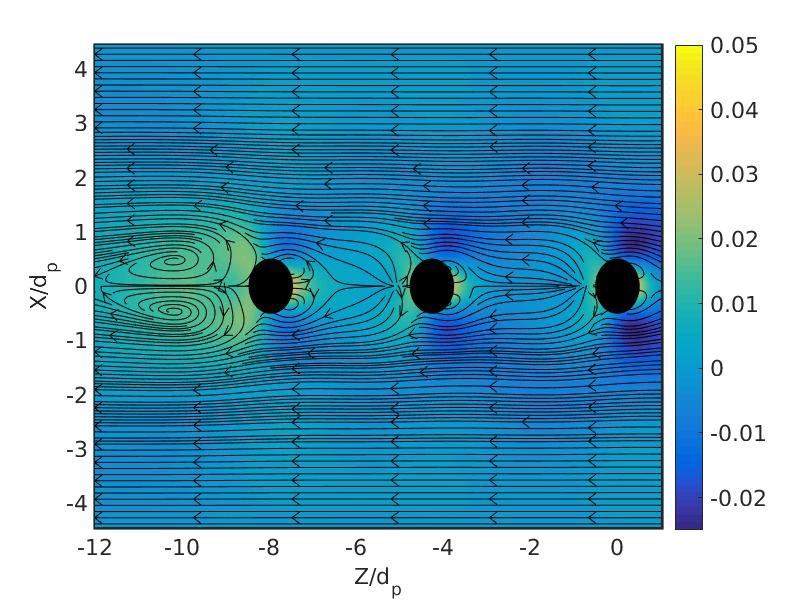}}
\put( -150,110) { \colorbox{white}{ \textcolor{black}{\bf (c)} } }
\end{center}
\vspace{-0.5cm}
\caption{Plots of velocity streamlines relative to the front particle in the microchannel in XZ plane for (a) a single, (b) two-particle train and (c) three-particle train at $\ReP = 1.5$. Colored contours show the streamwise velocity perturbation (subtracting the unperturbed channel flow from the instantaneous velocity field) scaled by the average flow velocity. Flow direction is from left to right. }\label{fig:pert_XZ}
\end{figure*}

From these results, it seems that the particle assembly under finite inertia is a weakly coupled system. The origin of particle alignment seems to result from a favorable vortex connection between consecutive particles as illustrated in figure \ref{fig:Pair_dyn}. The vortex generated behind the front particle interacts with the vortex induced in front of the lagging particle, minimizing by that the fluctuating kinetic energy (in a similar way to particles interacting in oscillatory fluid flows \cite{klotsa2007interaction}, as discussed by \cite{hood2017pairwise}). However the vortex connection does not seem to occur when the train exceeds a certain number of aligned particles. Velocity perturbation induced by the individual particle Stresslet in bounded shear flow decays as $1/r^2$ at a distance from the particle center $r<H$. Since the train morphology does not fundamentally change when the number of particles increases, i.e. inter-particle spacing does not decrease significantly when the train length increases, hydrodynamic repulsion between pairs cannot be the driving mechanism. Visualizations of the flow perturbation at the channel scale (figure \ref{fig:pert_XZ}) reveal that the assembled particles move like a unique coherent structure, with a perturbed outer region that expands as the number of aligned particles increases. 
%Nevertheless, in stable assemblies the distance between the front and second particle is larger than the distance between the second and third, etc.  
Sequences of snapshots for the velocity perturbations reveal that the vortex connection starts first between the lagging and the second last particle, etc... until reaching the front of the train. It is likely that the hydrodynamic perturbation induced by the large structure when its length reaches the channel scale, prohibits the vortex connection between the leading particle and the second one, pushing the front particle to move forward. \\

%%%%%%%%%%%%%%% Perturbation %%%%%%%%%%%%%%%%%%%%%%%%%%%

Nevertheless, the observed departure of the leading particle, for instance in the simulations corresponding to figure \ref{fig:four_part}, does not depend drastically on the accuracy of the computed hydrodynamic perturbation induced by each particle. Since the Stresslet terms are the essential ingredients to capture the hydrodynamic interaction of the neutrally buoyant particles with the shear flow, we tuned the Stresslets terms in order to test whether this has an impact on the train stability. When we realized simulations with constant-value Stresslets (obtained from the converged three-particle train), instead of updating the Stresslets to maintain the zero-average strain rate inside the particle volume (eq. \ref{eq:FCM5}), very similar relative trajectories were observed. This suggests that the conditional stability described in this paper is a robust phenomenon that does not exclusively depend on the accuracy of the interactions at small (particle) scales. \\

\begin{figure}[h!]
\hspace{-0.5cm}
\begin{center}
{\includegraphics[width=0.8\linewidth]{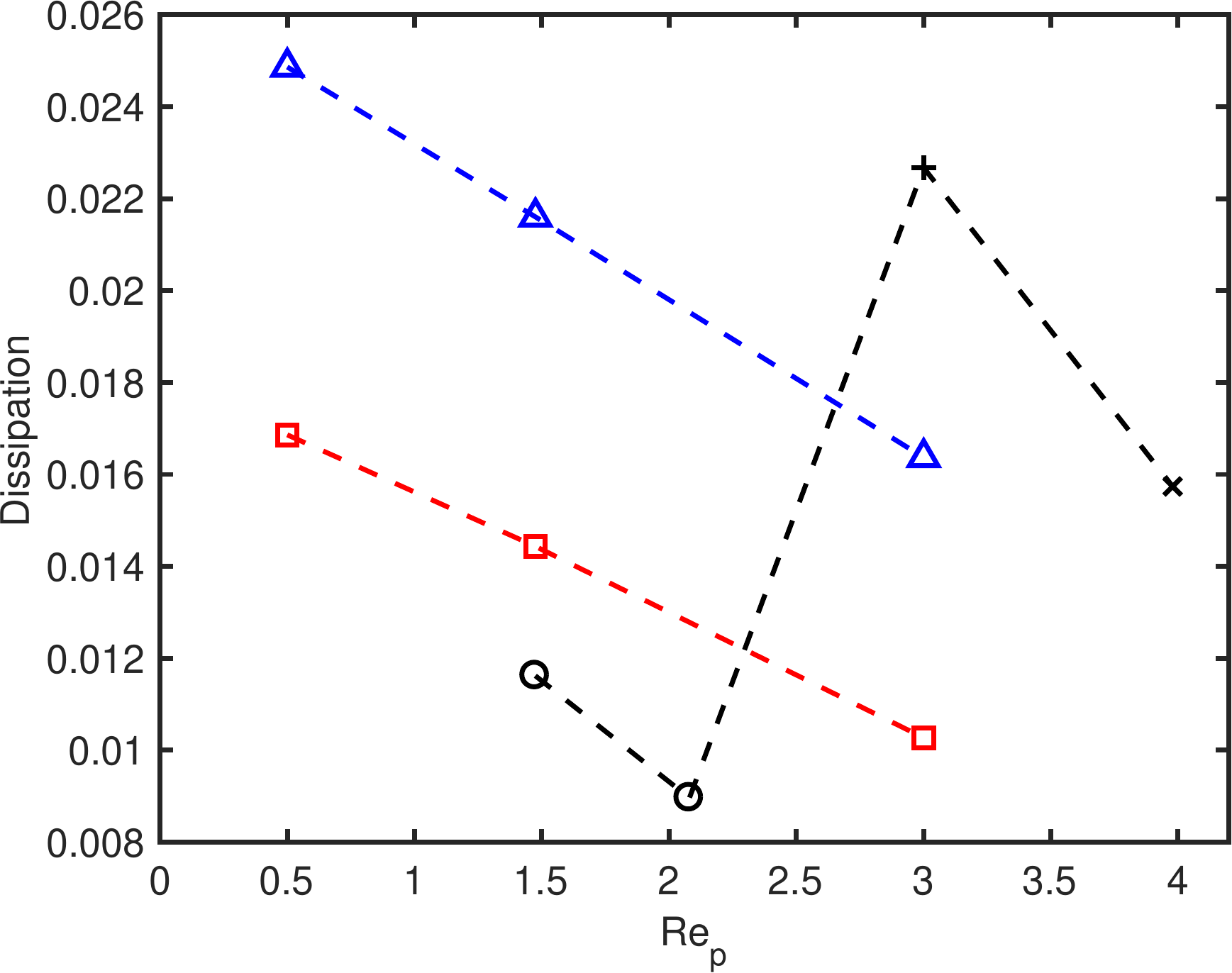}}
\end{center}
\vspace{-0.6cm}
\caption{The ratio between the drag force on particle train and forward pressure force applied by the fluid on the front particle. The symbol legend is idem to figure \ref{fig:train_stat}. }\label{fig:dissipation_ratio}
\end{figure}

A stability analysis, hard to realize on this system, would probably help to rationalize the effect of increasing the particle number on the stability of particle alignment. Here we limit our argument to the energy budget. At a given Reynolds number, when the number of particles assembled in a train increases, the slip velocity between the train and the ambient fluid flow increases. Although the neutrally buoyant particles move force-free in the flow, the slip velocity induces energy dissipation. The ratio between the energy dissipated by the train and the energy of the flow pushing forward the leading particle is plotted in figure \ref{fig:dissipation_ratio}. This ratio increases with the particle number and confinement, and it decreases with the Reynolds number. Figure \ref{fig:dissipation_ratio} suggests that the particle assembly should not cost above a threshold (around $2.5\%$ of the flow energy on the particle scale) for the system to remain stable.    \\

%%%%
\section{6. Concluding remarks }

After the validation of the numerical method, we gave some insight on the dynamics of a pair of neutrally buoyant particles that tend to align in the streamwise direction and form trains in channel flows. All the results were obtained after the inertial migration stages were accomplished, where the particles were located close to stable equilibrium spots and in one shear plane. The simulations were realized in domains long enough to eliminate any influence from the streamwise periodic boundaries. Trains of particles revealed to be slightly closer to the channel walls than a single particle at equilibrium, and therefore they had a slower streamwise velocity. The trains were slightly inclined with respect to the flow direction (lifted forward) when the Reynolds number increases, as already observed in experiments realized in tube flow \citep{matas2004trains}. The trains were unconditionally stable only in a limited range of Reynolds numbers and particle diameter -to- channel height ratios. When the train length increases, the hydrodynamic perturbation induced by the train structure, is likely stronger than the perturbation induced by the Stresslets at the individual particle scale, pushing . \\

Our numerical results, obtained using a truncated multipole expansion, agree qualitatively well with the experiments \cite{gao2017self} realized within the same range of operating conditions. Future investigation on the interaction between one and several pairs of spheres near channel walls are required to assess quantitatively the bifurcation between stable and unstable alignment. Moreover, the conclusions on the particle assembly are valid for moderate particle size and when the solid volumetric concentration is low. When the particle size is almost half of the channel height, additional sets of equilibrium positions take place alternating on opposite walls, like in \cite{humphry2010axial, kahkeshani2016preferred}. This situation could not be examined by the Force Coupling Method as implemented here, mainly because of the truncation of the multipole expansion used in equations $[\ref{eq:FCM1} - \ref{eq:FCM3}]$. When the suspension volumetric concentration is not negligible ($\phi \succsim 0.5\%$) hydrodynamic dispersion is expected to decrease the alignment potentiality in a way complex to predict.  

%%%%%%%%%%%
\section{Acknowledgement}
%%%%%%%%%%%

This work was granted access to the HPC resources of CALMIP under the allocation 2017-P1002 and of GENCI under the allocations x20162b6942 and A0012B06942.

%%%%%%%%%%%%%%%%%%%%%%%%%%%%%%%%%%%%%%%%%%%%%%%%%%%%%%%%%%%%%%%%%%%%%%%%
\section{Appendix A: Particle near a wall in a linear flow}\label{lift_s}

The validation of the wall-normal force calculation was realized first by placing a neutrally-buoyant particle of radius $a$ near the bottom wall of a plane Couette flow. The domain size was $10.6a$ in the flow $x$ and wall-normal $y$-directions, and $8.1a$ in the spanwise $z$-direction. The computational grid was uniform, and the mesh size was $\Delta x=a/4$. Periodic boundary conditions were used in the $x$ and $z$ directions. The bottom wall was stationary and the top wall moving with $\VW = \dot \gamma H$, where $H$ is the distance between the walls and  $\dot \gamma$ is the shear rate. The particle was placed at a given position, and iterations were realized to find the force required to prohibit particle motion in the wall normal direction. \\

\begin{figure} [h]
\includegraphics[width=0.49\linewidth]{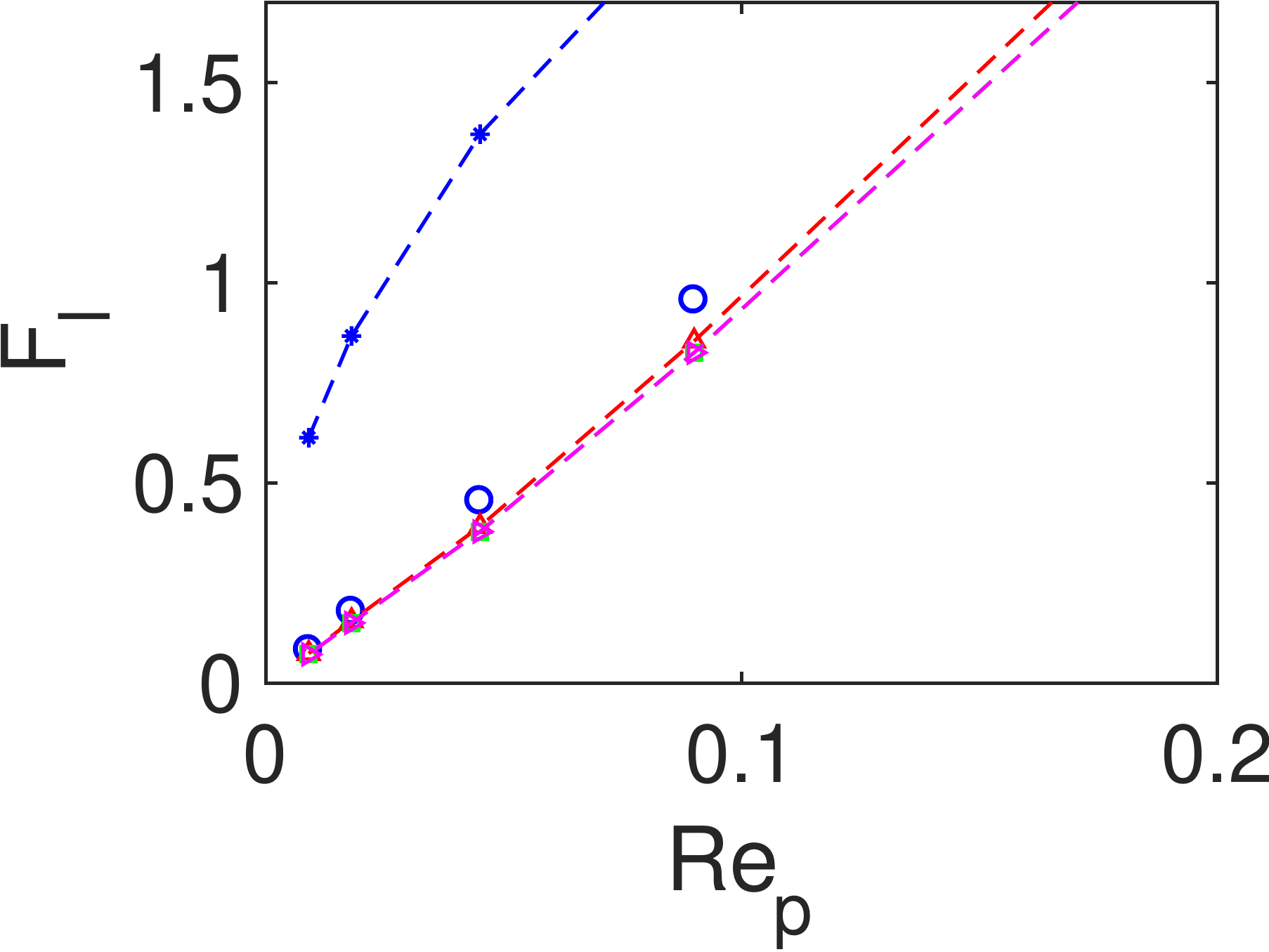}
\includegraphics[width=0.49\linewidth]{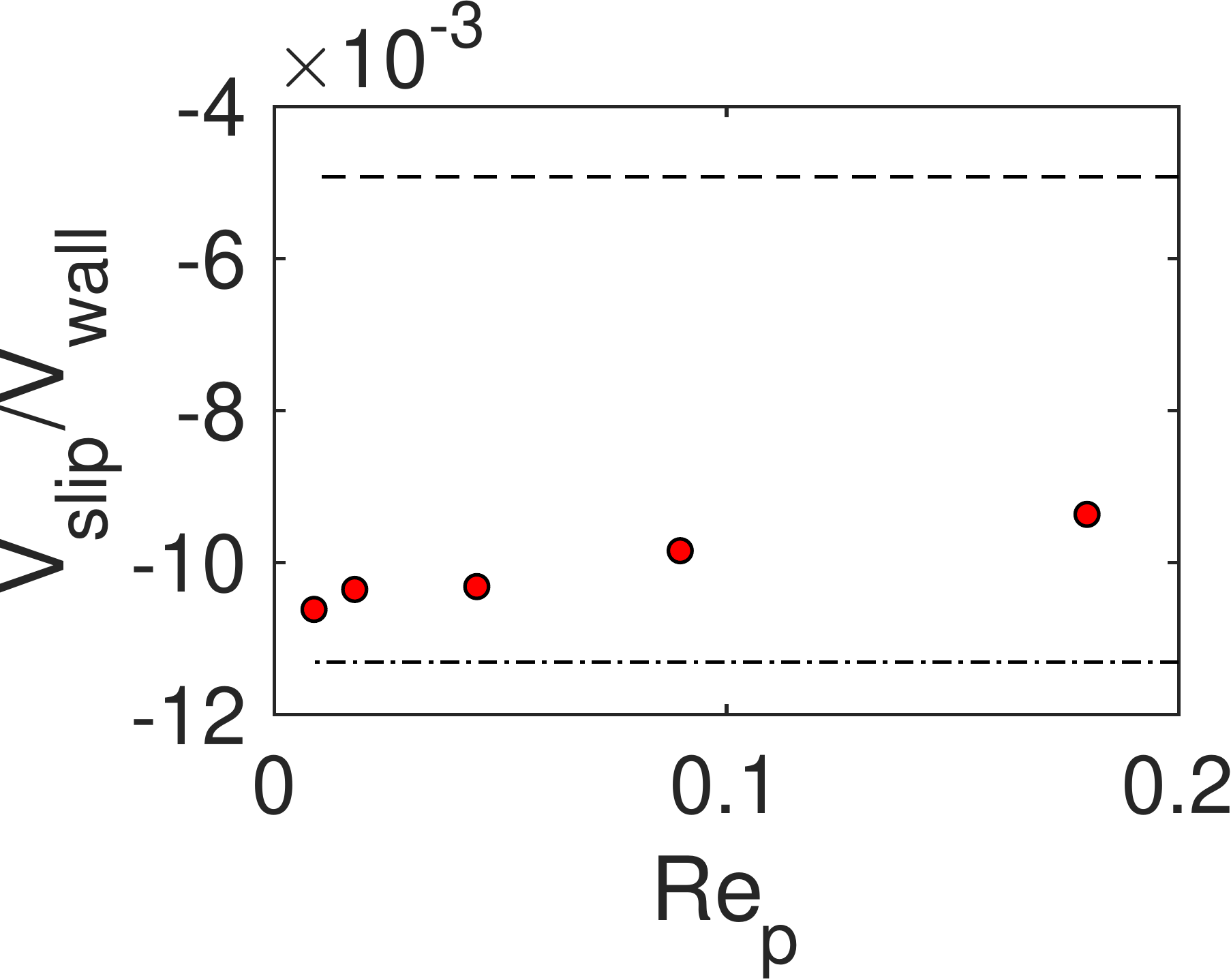}
\caption{Left Panel: Numerical calculation of the wall-normal force (blue open circles) applied on a particle near a wall in linear flow, and comparison with different theoretical predictions. The force is scaled by $\mu a V_{slip}$. The different theories are from Saffman~\cite{saffman1965lift} (blue asteriks), Cherukat \& Mclaughlin~\cite{cherukat1994inertial} (red upward-pointing triangle), Lovalenti in the appendix of \cite{cherukat1994inertial} (green square) and Magnaudet~\cite{magnaudet2003small} (magenta right-pointing triangle). Rigth Panel: streamwise particle slip velocity versus $Re_p$. The red circles are from FCM simulations. The lines are the prediction of the slip velocity from \citet{goldman1967slow} at $y_p=1.54$ and $2.35a$ respectively.}\label{fig:Fl}
\end{figure}

An example of the results is shown in figure \subfig{fig:Fl}{a} obtained by placing the particle at $\yPi = 1.66a$ near the bottom wall. This figure shows the increase of the wall-normal force with the Reynolds number defined in linear flow as $\ReP=\dot \gamma a^2/\nu$. The force is scaled by the viscous drag, $\mu a V_{slip}$, where $V_{slip}$ is the particle slip velocity (in the streamwise direction) with respect to the unperturbed local fluid flow, and $\mu$ is the dynamic fluid viscosity. Note that $V_{slip}$ is not known $a ~ priori$, but calculated from the simulation result at equilibrium, upon completion of the iterative procedure used to obtain the force. \subfig{fig:Fl}{b} shows $V_{slip}$ scaled by the wall velocity as a function of $Re_p$.  It is compared in the same graph with the Stokes flow limit from ~\cite{goldman1967slow} at the two wall-normal positions $y_p=1.54$ and $2.35a$ reported in Table 2 of their paper. Note that in our simulations the particle position $y_p$, initially equal to $1.5a$ in the simulations at different $Re_p$ is found between $1.66a$ and $1.7a$ at the end of the iterative procedure. The calculated slip is close to the Stokes flow prediction. Its amplitude decreases slightly with the Reynolds number. This is not an inertial effect. It is rather related to the fact that the steady particle position is not the same. \\

Several theoretical works allowed determining the lift force applied on a particle in a linear flow assuming no fluid acceleration at the particle scale. In this paper, we compare the numerical results with that of ~\cite{saffman1965lift, cherukat1994inertial, magnaudet2003small}. The expressions of the lift force obtained by the different works are listed below. All of them are scaled with $\mu a V_{slip}$ and take into account the proximity of the particle to the wall, except the expression of Saffman \cite{saffman1965lift} obtained in unbounded shear flow. Figure \ref{fig:Fl} shows that the numerical results agree very well with the theories that take into account the wall presence.

The expressions of the lift force resulting on a finite size particle in a linear flow are given in this appendix, from different sources in the literature. The first one does not take into account the presence of the wall. The last three contain the parameter $\kappa$ which is the ratio between the particle position with respect to the closest wall and the particle radius:

\begin{itemize}
\item
Eq. (3.11) of Saffman 1965 ~\cite{saffman1965lift}, with a correction of $\frac{1}{4\pi}$ published in the erratum:
\begin{eqnarray}
F_l & = & \frac{81.2}{2 \pi} a \sqrt{\dot \gamma /\nu} = 6.46 \sqrt{\ReP}
\end{eqnarray}
\item
Eq. (4.2) of Cherukat and McLaughlin 1994 ~\cite{cherukat1994inertial}. This work also accounts for the distance between wall and particle; $\kappa = a/\yPi$ and valid in the regime $\yPi << min(\Ls,\Lg)$, where Stokes length $\Ls = \nu/V_{slip}$, Saffman length $\Lg = (\nu/\dot \gamma)^{1/2}$ and dimensionless parameter $\lamg = \dot \gamma a/V_{slip}$
\begin{center}
\begin{eqnarray}
F_l & = & \Res I = \ReP (A \lamg^{-1} - B + C \lamg)
\end{eqnarray}
\end{center}
Where $\Res = a V_{slip}/\nu$; $A = 1.7631 + 0.3561 \kappa - 1.1837 \kappa^2 + 0.84516 \kappa^3$, $B = 3.24139/\kappa + 2.676 + 0.8248 \kappa - 0.4616 \kappa^2$ and $C = 1.8081 + 0.879585 \kappa - 1.9009 \kappa^2 + 0.98149 \kappa^3$.
\item 
Eq. (A 14) of Cherukat and McLaughlin 1994 ~\cite{cherukat1994inertial}, which is derived by Lovalenti. We non-dimensionalized eq (A 14) by $(\rho \nu a V_{slip})$.

\begin{equation}
\begin{split}
F_l & = \frac{18 \pi}{32} \Res +\frac 7 8 \pi \ReP  \\
& -\frac{66 \pi}{ 64} \ReP \left( \frac1\kappa + \frac{27}{16} \right) \left( \frac{6 \pi 55}{144 \times 4} \right) \ReP \left( \frac{a \dot \gamma}{V_{slip}}\right) \nonumber \\
&      = \pi \ReP \\
& \left [ \frac{18}{32} \frac 1 \lamg +\frac 7 8 -\frac{66}{64} \left( \frac1\kappa + \frac{27}{16} \right) \left( \frac{6 \times 55}{144 \times 4} \right) \lamg \right ] 
\end{split}
\end{equation}

\item Eq. (17) of Magnaudet 2003 ~\cite{magnaudet2003small}, which is valid for spherical bubble, droplet and particle. It is valid for the object close to the wall. For particles, in case of viscosity ratio $\lambda \to \infty$, the non-dimensionalized lift force is:

\begin{equation}
\begin{split}
F_l & = \frac{\pi}{4} \Res (3/2)^2 \left[ (1+ 11/8 \lamg^2 5/3) - \frac{11}{6} \lamg (\kappa^{-1} + 0.84) \right]  \nonumber \\
& = \frac{9 \pi}{16} \ReP \left[ (\lamg^{-1}+ 55/54 \lamg) - \frac{11}{6} (\kappa^{-1} + 0.84) \right]
\end{split}
\end{equation}

\end{itemize}

 \bibliography{Biblio}

\begin{thebibliography}{40}%
\makeatletter
\providecommand \@ifxundefined [1]{%
 \@ifx{#1\undefined}
}%
\providecommand \@ifnum [1]{%
 \ifnum #1\expandafter \@firstoftwo
 \else \expandafter \@secondoftwo
 \fi
}%
\providecommand \@ifx [1]{%
 \ifx #1\expandafter \@firstoftwo
 \else \expandafter \@secondoftwo
 \fi
}%
\providecommand \natexlab [1]{#1}%
\providecommand \enquote  [1]{``#1''}%
\providecommand \bibnamefont  [1]{#1}%
\providecommand \bibfnamefont [1]{#1}%
\providecommand \citenamefont [1]{#1}%
\providecommand \href@noop [0]{\@secondoftwo}%
\providecommand \href [0]{\begingroup \@sanitize@url \@href}%
\providecommand \@href[1]{\@@startlink{#1}\@@href}%
\providecommand \@@href[1]{\endgroup#1\@@endlink}%
\providecommand \@sanitize@url [0]{\catcode `\\12\catcode `\$12\catcode
  `\&12\catcode `\#12\catcode `\^12\catcode `\_12\catcode `\%12\relax}%
\providecommand \@@startlink[1]{}%
\providecommand \@@endlink[0]{}%
\providecommand \url  [0]{\begingroup\@sanitize@url \@url }%
\providecommand \@url [1]{\endgroup\@href {#1}{\urlprefix }}%
\providecommand \urlprefix  [0]{URL }%
\providecommand \Eprint [0]{\href }%
\providecommand \doibase [0]{http://dx.doi.org/}%
\providecommand \selectlanguage [0]{\@gobble}%
\providecommand \bibinfo  [0]{\@secondoftwo}%
\providecommand \bibfield  [0]{\@secondoftwo}%
\providecommand \translation [1]{[#1]}%
\providecommand \BibitemOpen [0]{}%
\providecommand \bibitemStop [0]{}%
\providecommand \bibitemNoStop [0]{.\EOS\space}%
\providecommand \EOS [0]{\spacefactor3000\relax}%
\providecommand \BibitemShut  [1]{\csname bibitem#1\endcsname}%
\let\auto@bib@innerbib\@empty
%</preamble>
\bibitem [{\citenamefont {Segr{\'e}}\ and\ \citenamefont
  {Silberberg}(1962)}]{segre1962behaviour}%
  \BibitemOpen
  \bibfield  {author} {\bibinfo {author} {\bibfnamefont {G}~\bibnamefont
  {Segr{\'e}}}\ and\ \bibinfo {author} {\bibfnamefont {A}~\bibnamefont
  {Silberberg}},\ }\bibfield  {title} {\enquote {\bibinfo {title} {Behaviour of
  macroscopic rigid spheres in poiseuille flow part 2. experimental results and
  interpretation},}\ }\href@noop {} {\bibfield  {journal} {\bibinfo  {journal}
  {Journal of fluid mechanics}\ }\textbf {\bibinfo {volume} {14}},\ \bibinfo
  {pages} {136--157} (\bibinfo {year} {1962})}\BibitemShut {NoStop}%
\bibitem [{\citenamefont {Ho}\ and\ \citenamefont
  {Leal}(1974)}]{ho1974inertial}%
  \BibitemOpen
  \bibfield  {author} {\bibinfo {author} {\bibfnamefont {BP}~\bibnamefont
  {Ho}}\ and\ \bibinfo {author} {\bibfnamefont {LG}~\bibnamefont {Leal}},\
  }\bibfield  {title} {\enquote {\bibinfo {title} {Inertial migration of rigid
  spheres in two-dimensional unidirectional flows},}\ }\href@noop {} {\bibfield
   {journal} {\bibinfo  {journal} {Journal of fluid mechanics}\ }\textbf
  {\bibinfo {volume} {65}},\ \bibinfo {pages} {365--400} (\bibinfo {year}
  {1974})}\BibitemShut {NoStop}%
\bibitem [{\citenamefont {Vasseur}\ and\ \citenamefont
  {Cox}(1976)}]{vasseur1976lateral}%
  \BibitemOpen
  \bibfield  {author} {\bibinfo {author} {\bibfnamefont {P}~\bibnamefont
  {Vasseur}}\ and\ \bibinfo {author} {\bibfnamefont {RG}~\bibnamefont {Cox}},\
  }\bibfield  {title} {\enquote {\bibinfo {title} {The lateral migration of a
  spherical particle in two-dimensional shear flows},}\ }\href@noop {}
  {\bibfield  {journal} {\bibinfo  {journal} {Journal of Fluid Mechanics}\
  }\textbf {\bibinfo {volume} {78}},\ \bibinfo {pages} {385--413} (\bibinfo
  {year} {1976})}\BibitemShut {NoStop}%
\bibitem [{\citenamefont {Schonberg}\ and\ \citenamefont
  {Hinch}(1989)}]{schonberg1989inertial}%
  \BibitemOpen
  \bibfield  {author} {\bibinfo {author} {\bibfnamefont {J.~A}\ \bibnamefont
  {Schonberg}}\ and\ \bibinfo {author} {\bibfnamefont {EJ}~\bibnamefont
  {Hinch}},\ }\bibfield  {title} {\enquote {\bibinfo {title} {Inertial
  migration of a sphere in poiseuille flow},}\ }\href@noop {} {\bibfield
  {journal} {\bibinfo  {journal} {Journal of Fluid Mechanics}\ }\textbf
  {\bibinfo {volume} {203}},\ \bibinfo {pages} {517--524} (\bibinfo {year}
  {1989})}\BibitemShut {NoStop}%
\bibitem [{\citenamefont {Asmolov}(1999)}]{asmolov1999inertial}%
  \BibitemOpen
  \bibfield  {author} {\bibinfo {author} {\bibfnamefont {E.S.}\ \bibnamefont
  {Asmolov}},\ }\bibfield  {title} {\enquote {\bibinfo {title} {The inertial
  lift on a spherical particle in a plane poiseuille flow at large channel
  reynolds number},}\ }\href@noop {} {\bibfield  {journal} {\bibinfo  {journal}
  {Journal of Fluid Mechanics}\ }\textbf {\bibinfo {volume} {381}},\ \bibinfo
  {pages} {63--87} (\bibinfo {year} {1999})}\BibitemShut {NoStop}%
\bibitem [{\citenamefont {Oakey}\ \emph {et~al.}(2010)\citenamefont {Oakey},
  \citenamefont {J.R.W.}, \citenamefont {Arellano}, \citenamefont {Di~Carlo},
  \citenamefont {Graves},\ and\ \citenamefont {Toner}}]{oakey2010particle}%
  \BibitemOpen
  \bibfield  {author} {\bibinfo {author} {\bibfnamefont {J.}~\bibnamefont
  {Oakey}}, \bibinfo {author} {\bibfnamefont {Applegate}\ \bibnamefont
  {J.R.W.}}, \bibinfo {author} {\bibfnamefont {E.}~\bibnamefont {Arellano}},
  \bibinfo {author} {\bibfnamefont {D.}~\bibnamefont {Di~Carlo}}, \bibinfo
  {author} {\bibfnamefont {S.~W}\ \bibnamefont {Graves}}, \ and\ \bibinfo
  {author} {\bibfnamefont {M.}~\bibnamefont {Toner}},\ }\bibfield  {title}
  {\enquote {\bibinfo {title} {Particle focusing in staged inertial
  microfluidic devices for flow cytometry},}\ }\href@noop {} {\bibfield
  {journal} {\bibinfo  {journal} {Analytical chemistry}\ }\textbf {\bibinfo
  {volume} {82}},\ \bibinfo {pages} {3862--3867} (\bibinfo {year}
  {2010})}\BibitemShut {NoStop}%
\bibitem [{\citenamefont {Edd}\ \emph {et~al.}(2008)\citenamefont {Edd},
  \citenamefont {Di~Carlo}, \citenamefont {Humphry}, \citenamefont
  {K{\"o}ster}, \citenamefont {Irimia}, \citenamefont {Weitz},\ and\
  \citenamefont {Toner}}]{edd2008controlled}%
  \BibitemOpen
  \bibfield  {author} {\bibinfo {author} {\bibfnamefont {J.F}\ \bibnamefont
  {Edd}}, \bibinfo {author} {\bibfnamefont {D.}~\bibnamefont {Di~Carlo}},
  \bibinfo {author} {\bibfnamefont {K.J.}\ \bibnamefont {Humphry}}, \bibinfo
  {author} {\bibfnamefont {S.}~\bibnamefont {K{\"o}ster}}, \bibinfo {author}
  {\bibfnamefont {D.}~\bibnamefont {Irimia}}, \bibinfo {author} {\bibfnamefont
  {D.~A}\ \bibnamefont {Weitz}}, \ and\ \bibinfo {author} {\bibfnamefont
  {M.}~\bibnamefont {Toner}},\ }\bibfield  {title} {\enquote {\bibinfo {title}
  {Controlled encapsulation of single-cells into monodisperse picolitre
  drops},}\ }\href@noop {} {\bibfield  {journal} {\bibinfo  {journal} {Lab on a
  Chip}\ }\textbf {\bibinfo {volume} {8}},\ \bibinfo {pages} {1262--1264}
  (\bibinfo {year} {2008})}\BibitemShut {NoStop}%
\bibitem [{\citenamefont {Hur}\ \emph {et~al.}(2010)\citenamefont {Hur},
  \citenamefont {Tse},\ and\ \citenamefont {Di~Carlo}}]{hur2010sheathless}%
  \BibitemOpen
  \bibfield  {author} {\bibinfo {author} {\bibfnamefont {S.C.}\ \bibnamefont
  {Hur}}, \bibinfo {author} {\bibfnamefont {H.~Tat~K.}\ \bibnamefont {Tse}}, \
  and\ \bibinfo {author} {\bibfnamefont {D.}~\bibnamefont {Di~Carlo}},\
  }\bibfield  {title} {\enquote {\bibinfo {title} {Sheathless inertial cell
  ordering for extreme throughput flow cytometry},}\ }\href@noop {} {\bibfield
  {journal} {\bibinfo  {journal} {Lab on a Chip}\ }\textbf {\bibinfo {volume}
  {10}},\ \bibinfo {pages} {274--280} (\bibinfo {year} {2010})}\BibitemShut
  {NoStop}%
\bibitem [{\citenamefont {Matas}\ \emph
  {et~al.}(2004{\natexlab{a}})\citenamefont {Matas}, \citenamefont {Morris},\
  and\ \citenamefont {Guazzelli}}]{matas2004}%
  \BibitemOpen
  \bibfield  {author} {\bibinfo {author} {\bibfnamefont {J.-P.}\ \bibnamefont
  {Matas}}, \bibinfo {author} {\bibfnamefont {J.~F.}\ \bibnamefont {Morris}}, \
  and\ \bibinfo {author} {\bibfnamefont {{\'E}}~\bibnamefont {Guazzelli}},\
  }\bibfield  {title} {\enquote {\bibinfo {title} {Inertial migration of rigid
  spherical particles in poiseuille flow},}\ }\href@noop {} {\bibfield
  {journal} {\bibinfo  {journal} {Journal of fluid mechanics}\ }\textbf
  {\bibinfo {volume} {515}},\ \bibinfo {pages} {171–195} (\bibinfo {year}
  {2004}{\natexlab{a}})}\BibitemShut {NoStop}%
\bibitem [{\citenamefont {Di~Carlo}\ \emph {et~al.}(2007)\citenamefont
  {Di~Carlo}, \citenamefont {Irimia}, \citenamefont {Tompkins},\ and\
  \citenamefont {Toner}}]{di2007continuous}%
  \BibitemOpen
  \bibfield  {author} {\bibinfo {author} {\bibfnamefont {D}~\bibnamefont
  {Di~Carlo}}, \bibinfo {author} {\bibfnamefont {D}~\bibnamefont {Irimia}},
  \bibinfo {author} {\bibfnamefont {R~G}\ \bibnamefont {Tompkins}}, \ and\
  \bibinfo {author} {\bibfnamefont {M}~\bibnamefont {Toner}},\ }\bibfield
  {title} {\enquote {\bibinfo {title} {Continuous inertial focusing, ordering,
  and separation of particles in microchannels},}\ }\href@noop {} {\bibfield
  {journal} {\bibinfo  {journal} {Proceedings of the National Academy of
  Sciences}\ }\textbf {\bibinfo {volume} {104}},\ \bibinfo {pages}
  {18892--18897} (\bibinfo {year} {2007})}\BibitemShut {NoStop}%
\bibitem [{\citenamefont {Bhagat}\ \emph {et~al.}(2008)\citenamefont {Bhagat},
  \citenamefont {Kuntaegowdanahalli},\ and\ \citenamefont
  {Papautsky}}]{bhagat2008enhanced}%
  \BibitemOpen
  \bibfield  {author} {\bibinfo {author} {\bibfnamefont {A.A.S}\ \bibnamefont
  {Bhagat}}, \bibinfo {author} {\bibfnamefont {S.S.}\ \bibnamefont
  {Kuntaegowdanahalli}}, \ and\ \bibinfo {author} {\bibfnamefont
  {I.}~\bibnamefont {Papautsky}},\ }\bibfield  {title} {\enquote {\bibinfo
  {title} {Enhanced particle filtration in straight microchannels using
  shear-modulated inertial migration},}\ }\href@noop {} {\bibfield  {journal}
  {\bibinfo  {journal} {Physics of Fluids}\ }\textbf {\bibinfo {volume} {20}},\
  \bibinfo {pages} {101702} (\bibinfo {year} {2008})}\BibitemShut {NoStop}%
\bibitem [{\citenamefont {Choi}\ \emph {et~al.}(2011)\citenamefont {Choi},
  \citenamefont {Seo},\ and\ \citenamefont {Lee}}]{choi2011lateral}%
  \BibitemOpen
  \bibfield  {author} {\bibinfo {author} {\bibfnamefont {Y.S}\ \bibnamefont
  {Choi}}, \bibinfo {author} {\bibfnamefont {K.W.}\ \bibnamefont {Seo}}, \ and\
  \bibinfo {author} {\bibfnamefont {S.J}\ \bibnamefont {Lee}},\ }\bibfield
  {title} {\enquote {\bibinfo {title} {Lateral and cross-lateral focusing of
  spherical particles in a square microchannel},}\ }\href@noop {} {\bibfield
  {journal} {\bibinfo  {journal} {Lab on a Chip}\ }\textbf {\bibinfo {volume}
  {11}},\ \bibinfo {pages} {460--465} (\bibinfo {year} {2011})}\BibitemShut
  {NoStop}%
\bibitem [{\citenamefont {Matas}\ \emph
  {et~al.}(2004{\natexlab{b}})\citenamefont {Matas}, \citenamefont {Glezer},
  \citenamefont {Guazzelli},\ and\ \citenamefont {Morris}}]{matas2004trains}%
  \BibitemOpen
  \bibfield  {author} {\bibinfo {author} {\bibfnamefont {J.P.}\ \bibnamefont
  {Matas}}, \bibinfo {author} {\bibfnamefont {V.}~\bibnamefont {Glezer}},
  \bibinfo {author} {\bibfnamefont {{\'E}}~\bibnamefont {Guazzelli}}, \ and\
  \bibinfo {author} {\bibfnamefont {J.F.}\ \bibnamefont {Morris}},\ }\bibfield
  {title} {\enquote {\bibinfo {title} {Trains of particles in
  finite-reynolds-number pipe flow},}\ }\href@noop {} {\bibfield  {journal}
  {\bibinfo  {journal} {Physics of Fluids}\ }\textbf {\bibinfo {volume} {16}},\
  \bibinfo {pages} {4192--4195} (\bibinfo {year}
  {2004}{\natexlab{b}})}\BibitemShut {NoStop}%
\bibitem [{\citenamefont {Chun}\ and\ \citenamefont
  {Ladd}(2006)}]{chun2006inertial}%
  \BibitemOpen
  \bibfield  {author} {\bibinfo {author} {\bibfnamefont {B}~\bibnamefont
  {Chun}}\ and\ \bibinfo {author} {\bibfnamefont {AJC}\ \bibnamefont {Ladd}},\
  }\bibfield  {title} {\enquote {\bibinfo {title} {Inertial migration of
  neutrally buoyant particles in a square duct: An investigation of multiple
  equilibrium positions},}\ }\href@noop {} {\bibfield  {journal} {\bibinfo
  {journal} {Physics of Fluids}\ }\textbf {\bibinfo {volume} {18}},\ \bibinfo
  {pages} {031704} (\bibinfo {year} {2006})}\BibitemShut {NoStop}%
\bibitem [{\citenamefont {Loisel}\ \emph {et~al.}(2015)\citenamefont {Loisel},
  \citenamefont {Abbas}, \citenamefont {Masbernat},\ and\ \citenamefont
  {Climent}}]{loisel2015inertia}%
  \BibitemOpen
  \bibfield  {author} {\bibinfo {author} {\bibfnamefont {V.}~\bibnamefont
  {Loisel}}, \bibinfo {author} {\bibfnamefont {M.}~\bibnamefont {Abbas}},
  \bibinfo {author} {\bibfnamefont {O.}~\bibnamefont {Masbernat}}, \ and\
  \bibinfo {author} {\bibfnamefont {E.}~\bibnamefont {Climent}},\ }\bibfield
  {title} {\enquote {\bibinfo {title} {Inertia-driven particle migration and
  mixing in a wall-bounded laminar suspension flow},}\ }\href@noop {}
  {\bibfield  {journal} {\bibinfo  {journal} {Physics of Fluids}\ }\textbf
  {\bibinfo {volume} {27}},\ \bibinfo {pages} {123304} (\bibinfo {year}
  {2015})}\BibitemShut {NoStop}%
\bibitem [{\citenamefont {Kahkeshani}\ \emph {et~al.}(2016)\citenamefont
  {Kahkeshani}, \citenamefont {Haddadi},\ and\ \citenamefont
  {Di~Carlo}}]{kahkeshani2016preferred}%
  \BibitemOpen
  \bibfield  {author} {\bibinfo {author} {\bibfnamefont {S.}~\bibnamefont
  {Kahkeshani}}, \bibinfo {author} {\bibfnamefont {H.}~\bibnamefont {Haddadi}},
  \ and\ \bibinfo {author} {\bibfnamefont {D.}~\bibnamefont {Di~Carlo}},\
  }\bibfield  {title} {\enquote {\bibinfo {title} {Preferred interparticle
  spacings in trains of particles in inertial microchannel flows},}\
  }\href@noop {} {\bibfield  {journal} {\bibinfo  {journal} {J. Fluid Mech}\
  }\textbf {\bibinfo {volume} {786}},\ \bibinfo {pages} {R3} (\bibinfo {year}
  {2016})}\BibitemShut {NoStop}%
\bibitem [{\citenamefont {Gao}\ \emph {et~al.}(2017)\citenamefont {Gao},
  \citenamefont {Magaud}, \citenamefont {Baldas}, \citenamefont {Lafforgue},
  \citenamefont {Abbas},\ and\ \citenamefont {Colin}}]{gao2017self}%
  \BibitemOpen
  \bibfield  {author} {\bibinfo {author} {\bibfnamefont {Y.}~\bibnamefont
  {Gao}}, \bibinfo {author} {\bibfnamefont {P.}~\bibnamefont {Magaud}},
  \bibinfo {author} {\bibfnamefont {L.}~\bibnamefont {Baldas}}, \bibinfo
  {author} {\bibfnamefont {C.}~\bibnamefont {Lafforgue}}, \bibinfo {author}
  {\bibfnamefont {M.}~\bibnamefont {Abbas}}, \ and\ \bibinfo {author}
  {\bibfnamefont {S.}~\bibnamefont {Colin}},\ }\bibfield  {title} {\enquote
  {\bibinfo {title} {Self-ordered particle trains in inertial microchannel
  flows},}\ }\href@noop {} {\bibfield  {journal} {\bibinfo  {journal}
  {Microfluidics and Nanofluidics}\ }\textbf {\bibinfo {volume} {21}},\
  \bibinfo {pages} {154} (\bibinfo {year} {2017})}\BibitemShut {NoStop}%
\bibitem [{\citenamefont {Poe}\ and\ \citenamefont
  {Acrivos}(1975)}]{poe1975closed}%
  \BibitemOpen
  \bibfield  {author} {\bibinfo {author} {\bibfnamefont {GG}~\bibnamefont
  {Poe}}\ and\ \bibinfo {author} {\bibfnamefont {Andreas}\ \bibnamefont
  {Acrivos}},\ }\bibfield  {title} {\enquote {\bibinfo {title}
  {Closed-streamline flows past rotating single cylinders and spheres: inertia
  effects},}\ }\href@noop {} {\bibfield  {journal} {\bibinfo  {journal}
  {Journal of Fluid Mechanics}\ }\textbf {\bibinfo {volume} {72}},\ \bibinfo
  {pages} {605--623} (\bibinfo {year} {1975})}\BibitemShut {NoStop}%
\bibitem [{\citenamefont {Yan}\ \emph {et~al.}(2007)\citenamefont {Yan},
  \citenamefont {Morris},\ and\ \citenamefont {Koplik}}]{yan2007hydrodynamic}%
  \BibitemOpen
  \bibfield  {author} {\bibinfo {author} {\bibfnamefont {Yiguang}\ \bibnamefont
  {Yan}}, \bibinfo {author} {\bibfnamefont {Jeffrey~F}\ \bibnamefont {Morris}},
  \ and\ \bibinfo {author} {\bibfnamefont {Joel}\ \bibnamefont {Koplik}},\
  }\bibfield  {title} {\enquote {\bibinfo {title} {Hydrodynamic interaction of
  two particles in confined linear shear flow at finite reynolds number},}\
  }\href@noop {} {\bibfield  {journal} {\bibinfo  {journal} {Physics of
  fluids}\ }\textbf {\bibinfo {volume} {19}},\ \bibinfo {pages} {113305}
  (\bibinfo {year} {2007})}\BibitemShut {NoStop}%
\bibitem [{\citenamefont {Lee}\ \emph {et~al.}(2010)\citenamefont {Lee},
  \citenamefont {Amini}, \citenamefont {Stone},\ and\ \citenamefont
  {Di~Carlo}}]{lee2010dynamic}%
  \BibitemOpen
  \bibfield  {author} {\bibinfo {author} {\bibfnamefont {W.}~\bibnamefont
  {Lee}}, \bibinfo {author} {\bibfnamefont {H.}~\bibnamefont {Amini}}, \bibinfo
  {author} {\bibfnamefont {H.~A}\ \bibnamefont {Stone}}, \ and\ \bibinfo
  {author} {\bibfnamefont {D.}~\bibnamefont {Di~Carlo}},\ }\bibfield  {title}
  {\enquote {\bibinfo {title} {Dynamic self-assembly and control of
  microfluidic particle crystals},}\ }\href@noop {} {\bibfield  {journal}
  {\bibinfo  {journal} {Proceedings of the National Academy of Sciences}\
  }\textbf {\bibinfo {volume} {107}},\ \bibinfo {pages} {22413--22418}
  (\bibinfo {year} {2010})}\BibitemShut {NoStop}%
\bibitem [{\citenamefont {Wang}\ \emph {et~al.}(2017)\citenamefont {Wang},
  \citenamefont {Abbas},\ and\ \citenamefont {Climent}}]{wang2017modulation}%
  \BibitemOpen
  \bibfield  {author} {\bibinfo {author} {\bibfnamefont {G.}~\bibnamefont
  {Wang}}, \bibinfo {author} {\bibfnamefont {M.}~\bibnamefont {Abbas}}, \ and\
  \bibinfo {author} {\bibfnamefont {E.}~\bibnamefont {Climent}},\ }\bibfield
  {title} {\enquote {\bibinfo {title} {Modulation of large-scale structures by
  neutrally buoyant and inertial finite-size particles in turbulent couette
  flow},}\ }\href@noop {} {\bibfield  {journal} {\bibinfo  {journal} {Phys.
  Rev. Fluids}\ } (\bibinfo {year} {2017})}\BibitemShut {NoStop}%
\bibitem [{\citenamefont {Calmet}\ and\ \citenamefont
  {Magnaudet}(1997)}]{calmet1997large}%
  \BibitemOpen
  \bibfield  {author} {\bibinfo {author} {\bibfnamefont {I.}~\bibnamefont
  {Calmet}}\ and\ \bibinfo {author} {\bibfnamefont {J.}~\bibnamefont
  {Magnaudet}},\ }\bibfield  {title} {\enquote {\bibinfo {title} {Large-eddy
  simulation of high-schmidt number mass transfer in a turbulent channel
  flow},}\ }\href@noop {} {\bibfield  {journal} {\bibinfo  {journal} {Physics
  of Fluids}\ }\textbf {\bibinfo {volume} {9}},\ \bibinfo {pages} {438--455}
  (\bibinfo {year} {1997})}\BibitemShut {NoStop}%
\bibitem [{\citenamefont {Maxey}\ and\ \citenamefont
  {Patel}(2001)}]{maxey2001localized}%
  \BibitemOpen
  \bibfield  {author} {\bibinfo {author} {\bibfnamefont {MR}~\bibnamefont
  {Maxey}}\ and\ \bibinfo {author} {\bibfnamefont {BK}~\bibnamefont {Patel}},\
  }\bibfield  {title} {\enquote {\bibinfo {title} {Localized force
  representations for particles sedimenting in stokes flow},}\ }\href@noop {}
  {\bibfield  {journal} {\bibinfo  {journal} {International journal of
  multiphase flow}\ }\textbf {\bibinfo {volume} {27}},\ \bibinfo {pages}
  {1603--1626} (\bibinfo {year} {2001})}\BibitemShut {NoStop}%
\bibitem [{\citenamefont {Lomholt}\ and\ \citenamefont
  {Maxey}(2003)}]{lomholt2003force}%
  \BibitemOpen
  \bibfield  {author} {\bibinfo {author} {\bibfnamefont {S}~\bibnamefont
  {Lomholt}}\ and\ \bibinfo {author} {\bibfnamefont {MR}~\bibnamefont
  {Maxey}},\ }\bibfield  {title} {\enquote {\bibinfo {title} {Force-coupling
  method for particulate two-phase flow: Stokes flow},}\ }\href@noop {}
  {\bibfield  {journal} {\bibinfo  {journal} {Journal of Computational
  Physics}\ }\textbf {\bibinfo {volume} {184}},\ \bibinfo {pages} {381--405}
  (\bibinfo {year} {2003})}\BibitemShut {NoStop}%
\bibitem [{\citenamefont {Yeo}\ and\ \citenamefont
  {Maxey}(2013)}]{yeo2013dynamics}%
  \BibitemOpen
  \bibfield  {author} {\bibinfo {author} {\bibfnamefont {KM}~\bibnamefont
  {Yeo}}\ and\ \bibinfo {author} {\bibfnamefont {MR}~\bibnamefont {Maxey}},\
  }\bibfield  {title} {\enquote {\bibinfo {title} {Dynamics and rheology of
  concentrated, finite-reynolds-number suspensions in a homogeneous shear
  flow},}\ }\href@noop {} {\bibfield  {journal} {\bibinfo  {journal} {Physics
  of Fluids}\ }\textbf {\bibinfo {volume} {25}},\ \bibinfo {pages} {053303}
  (\bibinfo {year} {2013})}\BibitemShut {NoStop}%
\bibitem [{\citenamefont {Yeo}\ \emph {et~al.}(2010)\citenamefont {Yeo},
  \citenamefont {Dong}, \citenamefont {Climent},\ and\ \citenamefont
  {Maxey}}]{yeo2010modulation}%
  \BibitemOpen
  \bibfield  {author} {\bibinfo {author} {\bibfnamefont {Kyongmin}\
  \bibnamefont {Yeo}}, \bibinfo {author} {\bibfnamefont {Suchuan}\ \bibnamefont
  {Dong}}, \bibinfo {author} {\bibfnamefont {Eric}\ \bibnamefont {Climent}}, \
  and\ \bibinfo {author} {\bibfnamefont {Martin~R}\ \bibnamefont {Maxey}},\
  }\bibfield  {title} {\enquote {\bibinfo {title} {Modulation of homogeneous
  turbulence seeded with finite size bubbles or particles},}\ }\href@noop {}
  {\bibfield  {journal} {\bibinfo  {journal} {International Journal of
  Multiphase Flow}\ }\textbf {\bibinfo {volume} {36}},\ \bibinfo {pages}
  {221--233} (\bibinfo {year} {2010})}\BibitemShut {NoStop}%
\bibitem [{\citenamefont {Mikulencak}\ and\ \citenamefont
  {Morris}(2004)}]{mikulencak2004stationary}%
  \BibitemOpen
  \bibfield  {author} {\bibinfo {author} {\bibfnamefont {Duane~R}\ \bibnamefont
  {Mikulencak}}\ and\ \bibinfo {author} {\bibfnamefont {Jeffrey~F}\
  \bibnamefont {Morris}},\ }\bibfield  {title} {\enquote {\bibinfo {title}
  {Stationary shear flow around fixed and free bodies at finite reynolds
  number},}\ }\href@noop {} {\bibfield  {journal} {\bibinfo  {journal} {Journal
  of Fluid Mechanics}\ }\textbf {\bibinfo {volume} {520}},\ \bibinfo {pages}
  {215--242} (\bibinfo {year} {2004})}\BibitemShut {NoStop}%
\bibitem [{\citenamefont {Haddadi}\ and\ \citenamefont
  {Morris}(2015)}]{Haddadi2015}%
  \BibitemOpen
  \bibfield  {author} {\bibinfo {author} {\bibfnamefont {H.}~\bibnamefont
  {Haddadi}}\ and\ \bibinfo {author} {\bibfnamefont {J.~F.}\ \bibnamefont
  {Morris}},\ }\bibfield  {title} {\enquote {\bibinfo {title} {Topology of
  pair-sphere trajectories in finite inertia suspension shear flow and its
  effects on microstructure and rheology},}\ }\href@noop {} {\bibfield
  {journal} {\bibinfo  {journal} {Physics of Fluids}\ }\textbf {\bibinfo
  {volume} {27}},\ \bibinfo {pages} {1--21} (\bibinfo {year}
  {2015})}\BibitemShut {NoStop}%
\bibitem [{\citenamefont {Goldman}\ \emph {et~al.}(1967)\citenamefont
  {Goldman}, \citenamefont {Cox},\ and\ \citenamefont
  {Brenner}}]{goldman1967slow}%
  \BibitemOpen
  \bibfield  {author} {\bibinfo {author} {\bibfnamefont {AJ}~\bibnamefont
  {Goldman}}, \bibinfo {author} {\bibfnamefont {RG}~\bibnamefont {Cox}}, \ and\
  \bibinfo {author} {\bibfnamefont {H}~\bibnamefont {Brenner}},\ }\bibfield
  {title} {\enquote {\bibinfo {title} {Slow viscous motion of a sphere parallel
  to a plane wall—ii couette flow},}\ }\href@noop {} {\bibfield  {journal}
  {\bibinfo  {journal} {Chemical engineering science}\ }\textbf {\bibinfo
  {volume} {22}},\ \bibinfo {pages} {653--660} (\bibinfo {year}
  {1967})}\BibitemShut {NoStop}%
\bibitem [{\citenamefont {Cherukat}\ and\ \citenamefont
  {Mclaughlin}(1994)}]{cherukat1994inertial}%
  \BibitemOpen
  \bibfield  {author} {\bibinfo {author} {\bibfnamefont {P.}~\bibnamefont
  {Cherukat}}\ and\ \bibinfo {author} {\bibfnamefont {J.B.}\ \bibnamefont
  {Mclaughlin}},\ }\bibfield  {title} {\enquote {\bibinfo {title} {The inertial
  lift on a rigid sphere in a linear shear flow field near a flat wall},}\
  }\href@noop {} {\bibfield  {journal} {\bibinfo  {journal} {Journal of Fluid
  Mechanics}\ }\textbf {\bibinfo {volume} {263}},\ \bibinfo {pages} {1--18}
  (\bibinfo {year} {1994})}\BibitemShut {NoStop}%
\bibitem [{\citenamefont {Asmolov}\ \emph {et~al.}(2018)\citenamefont
  {Asmolov}, \citenamefont {Dubov}, \citenamefont {Nizkaya}, \citenamefont
  {Harting},\ and\ \citenamefont {Vinogradova}}]{asmolov2018inertial}%
  \BibitemOpen
  \bibfield  {author} {\bibinfo {author} {\bibfnamefont {Evgeny~S}\
  \bibnamefont {Asmolov}}, \bibinfo {author} {\bibfnamefont {Alexander~L}\
  \bibnamefont {Dubov}}, \bibinfo {author} {\bibfnamefont {Tatiana~V}\
  \bibnamefont {Nizkaya}}, \bibinfo {author} {\bibfnamefont {Jens}\
  \bibnamefont {Harting}}, \ and\ \bibinfo {author} {\bibfnamefont {Olga~I}\
  \bibnamefont {Vinogradova}},\ }\bibfield  {title} {\enquote {\bibinfo {title}
  {Inertial focusing of finite-size particles in microchannels},}\ }\href@noop
  {} {\bibfield  {journal} {\bibinfo  {journal} {Journal of Fluid Mechanics}\
  }\textbf {\bibinfo {volume} {840}},\ \bibinfo {pages} {613--630} (\bibinfo
  {year} {2018})}\BibitemShut {NoStop}%
\bibitem [{\citenamefont {Hood}\ \emph {et~al.}(2015)\citenamefont {Hood},
  \citenamefont {Lee},\ and\ \citenamefont {Roper}}]{hood2015inertial}%
  \BibitemOpen
  \bibfield  {author} {\bibinfo {author} {\bibfnamefont {K}~\bibnamefont
  {Hood}}, \bibinfo {author} {\bibfnamefont {S}~\bibnamefont {Lee}}, \ and\
  \bibinfo {author} {\bibfnamefont {M}~\bibnamefont {Roper}},\ }\bibfield
  {title} {\enquote {\bibinfo {title} {Inertial migration of a rigid sphere in
  three-dimensional poiseuille flow},}\ }\href@noop {} {\bibfield  {journal}
  {\bibinfo  {journal} {Journal of Fluid Mechanics}\ }\textbf {\bibinfo
  {volume} {765}},\ \bibinfo {pages} {452--479} (\bibinfo {year}
  {2015})}\BibitemShut {NoStop}%
\bibitem [{\citenamefont {Abbas}\ \emph {et~al.}(2014)\citenamefont {Abbas},
  \citenamefont {Magaud}, \citenamefont {Gao},\ and\ \citenamefont
  {Geoffroy}}]{abbas2014migration}%
  \BibitemOpen
  \bibfield  {author} {\bibinfo {author} {\bibfnamefont {M}~\bibnamefont
  {Abbas}}, \bibinfo {author} {\bibfnamefont {P}~\bibnamefont {Magaud}},
  \bibinfo {author} {\bibfnamefont {Y}~\bibnamefont {Gao}}, \ and\ \bibinfo
  {author} {\bibfnamefont {S}~\bibnamefont {Geoffroy}},\ }\bibfield  {title}
  {\enquote {\bibinfo {title} {Migration of finite sized particles in a laminar
  square channel flow from low to high reynolds numbers},}\ }\href@noop {}
  {\bibfield  {journal} {\bibinfo  {journal} {Physics of Fluids}\ }\textbf
  {\bibinfo {volume} {26}},\ \bibinfo {pages} {123301} (\bibinfo {year}
  {2014})}\BibitemShut {NoStop}%
\bibitem [{\citenamefont {Subramanian}\ and\ \citenamefont
  {Koch}(2006)}]{Subramanian2006}%
  \BibitemOpen
  \bibfield  {author} {\bibinfo {author} {\bibfnamefont {G.}~\bibnamefont
  {Subramanian}}\ and\ \bibinfo {author} {\bibfnamefont {D.}~\bibnamefont
  {Koch}},\ }\bibfield  {title} {\enquote {\bibinfo {title} {Inertial effects
  on the transfer of heat or mass from neutrally buoyant spheres in a steady
  linear velocity field},}\ }\href@noop {} {\bibfield  {journal} {\bibinfo
  {journal} {Physics of Fluids}\ }\textbf {\bibinfo {volume} {18}},\ \bibinfo
  {pages} {1--18} (\bibinfo {year} {2006})}\BibitemShut {NoStop}%
\bibitem [{sup()}]{supmat}%
  \BibitemOpen
  \href@noop {} {\bibinfo  {journal} {See Supplemental Material at [URL will be
  inserted by publisher] for sequence of particle positions in the channel}\
  }\BibitemShut {NoStop}%
\bibitem [{\citenamefont {Klotsa}\ \emph {et~al.}(2007)\citenamefont {Klotsa},
  \citenamefont {Swift}, \citenamefont {Bowley},\ and\ \citenamefont
  {King}}]{klotsa2007interaction}%
  \BibitemOpen
\bibfield  {journal} {  }\bibfield  {author} {\bibinfo {author} {\bibfnamefont
  {D}~\bibnamefont {Klotsa}}, \bibinfo {author} {\bibfnamefont {Michael~R}\
  \bibnamefont {Swift}}, \bibinfo {author} {\bibfnamefont {RM}~\bibnamefont
  {Bowley}}, \ and\ \bibinfo {author} {\bibfnamefont {PJ}~\bibnamefont
  {King}},\ }\bibfield  {title} {\enquote {\bibinfo {title} {Interaction of
  spheres in oscillatory fluid flows},}\ }\href@noop {} {\bibfield  {journal}
  {\bibinfo  {journal} {Physical Review E}\ }\textbf {\bibinfo {volume} {76}},\
  \bibinfo {pages} {056314} (\bibinfo {year} {2007})}\BibitemShut {NoStop}%
\bibitem [{\citenamefont {Hood}\ and\ \citenamefont
  {Roper}(2017)}]{hood2017pairwise}%
  \BibitemOpen
  \bibfield  {author} {\bibinfo {author} {\bibfnamefont {K.}~\bibnamefont
  {Hood}}\ and\ \bibinfo {author} {\bibfnamefont {M.}~\bibnamefont {Roper}},\
  }\bibfield  {title} {\enquote {\bibinfo {title} {Pairwise interactions in
  inertially-driven one-dimensional microfluidic crystals},}\ }\href@noop {}
  {\bibfield  {journal} {\bibinfo  {journal} {arXiv preprint arXiv:1706.09992}\
  } (\bibinfo {year} {2017})}\BibitemShut {NoStop}%
\bibitem [{\citenamefont {Humphry}\ \emph {et~al.}(2010)\citenamefont
  {Humphry}, \citenamefont {Kulkarni}, \citenamefont {Weitz}, \citenamefont
  {Morris},\ and\ \citenamefont {Stone}}]{humphry2010axial}%
  \BibitemOpen
  \bibfield  {author} {\bibinfo {author} {\bibfnamefont {K.J.}\ \bibnamefont
  {Humphry}}, \bibinfo {author} {\bibfnamefont {P.~M}\ \bibnamefont
  {Kulkarni}}, \bibinfo {author} {\bibfnamefont {D.A}\ \bibnamefont {Weitz}},
  \bibinfo {author} {\bibfnamefont {J.F.}\ \bibnamefont {Morris}}, \ and\
  \bibinfo {author} {\bibfnamefont {H.A}\ \bibnamefont {Stone}},\ }\bibfield
  {title} {\enquote {\bibinfo {title} {Axial and lateral particle ordering in
  finite reynolds number channel flows},}\ }\href@noop {} {\bibfield  {journal}
  {\bibinfo  {journal} {Physics of Fluids}\ }\textbf {\bibinfo {volume} {22}},\
  \bibinfo {pages} {081703} (\bibinfo {year} {2010})}\BibitemShut {NoStop}%
\bibitem [{\citenamefont {Saffman}(1965)}]{saffman1965lift}%
  \BibitemOpen
  \bibfield  {author} {\bibinfo {author} {\bibfnamefont {PGT}\ \bibnamefont
  {Saffman}},\ }\bibfield  {title} {\enquote {\bibinfo {title} {The lift on a
  small sphere in a slow shear flow},}\ }\href@noop {} {\bibfield  {journal}
  {\bibinfo  {journal} {Journal of fluid mechanics}\ }\textbf {\bibinfo
  {volume} {22}},\ \bibinfo {pages} {385--400} (\bibinfo {year}
  {1965})}\BibitemShut {NoStop}%
\bibitem [{\citenamefont {Magnaudet}(2003)}]{magnaudet2003small}%
  \BibitemOpen
  \bibfield  {author} {\bibinfo {author} {\bibfnamefont {J.}~\bibnamefont
  {Magnaudet}},\ }\bibfield  {title} {\enquote {\bibinfo {title} {Small
  inertial effects on a spherical bubble, drop or particle moving near a wall
  in a time-dependent linear flow},}\ }\href@noop {} {\bibfield  {journal}
  {\bibinfo  {journal} {Journal of Fluid Mechanics}\ }\textbf {\bibinfo
  {volume} {485}},\ \bibinfo {pages} {115--142} (\bibinfo {year}
  {2003})}\BibitemShut {NoStop}%
\end{thebibliography}%
%\begin{thebibliography}{}
%\end{thebibliography}

\end{document}